\documentclass[aps,prd,superscriptaddress,twocolumn]{revtex4-1}
\usepackage{amssymb}
\usepackage{amsmath}
\usepackage{wallpaper}
\usepackage{bm}
\usepackage{multirow}
\usepackage[colorlinks,linkcolor=blue, urlcolor=blue, anchorcolor=blue, citecolor=blue]{hyperref}

\begin{document}

\title{Astrophysical constraints on nuclear EOSs and coupling constants in RMF models}

\author{Cheng-Jun Xia}
\affiliation{Center for Gravitation and Cosmology, College of Physical Science and Technology, Yangzhou University, Yangzhou 225009, China}

\author{Wen-Jie Xie}
\email{xiewenjie@ycu.edu.cn}
\affiliation{Department of Physics, Yuncheng University, Yuncheng 044000, China}
\affiliation{Shanxi Province Intelligent Optoelectronic Sensing Application Technology Innovation Center, Yuncheng University, Yuncheng 044000, China}
\affiliation{Guangxi Key Laboratory of Nuclear Physics and Nuclear Technology,
Guangxi Normal University, Guilin 541004, China}

\author{Mohemmedelnazier Bakhiet}
\affiliation{Center for Gravitation and Cosmology, College of Physical Science and Technology, Yangzhou University, Yangzhou 225009, China}

\date{\today}
\begin{abstract}
Utilizing various astrophysical constraints on neutron star structures, we carry out a Bayesian analysis on the density-dependent behaviors of coupling constants in RMF models as well as the nuclear matter properties at supranuclear densities. The effective nucleon interactions in the isoscalar-scalar, isoscalar-vector, and isovector-vector channels are considered, where the corresponding coupling constants ($\alpha_S, \alpha_V, \alpha_{TV}$) are fixed by dividing entire density range into three regions with six independent parameters. In this work we focus on constraining the density-dependent point-coupling constants at supranuclear densities, while the coupling constants at subsaturation densities are derived from the covariant density functional DD-ME2. For those consistent with astrophysical observations, the coupling constants generally decrease with density and approach to small positive values at large enough densities, which qualitatively agrees with various RMF models.  The posterior probability density functions and their correlations of the coupling constants and various nuclear matter properties are examined as well.  At $1\sigma$ level, the constrained coupling constants at density $1.5n_0$ ($2.5n_0$) are $\alpha_S = 3.1^{+0.1}_{-0.05} (1.55^{+0.85}_{-0.2}) \times 10^{-4} \mathrm{MeV}^{-2}$, $\alpha_V = 2.3^{+0.1}_{-0.0} (1.3^{+0.55}_{-0.1}) \times 10^{-4} \mathrm{MeV}^{-2}$, and $\alpha_{TV}  = 2.05^{+0}_{-0.4} (2.05^{+0}_{-0.5})\times 10^{-5} \mathrm{MeV}^{-2}$. At larger densities, we find the lower limit of $\alpha_{TV}$ is not well constrained, so that more extensive calculations with larger number of free parameters are necessary. Finally, the constrained pressure of neutron star matter and symmetry energy of nuclear matter at $1.5n_0$ ($2.5n_0$)  are $P=10^{+0.0}_{-0.0}$ ($60^{+15}_{-20}$) $\mathrm{MeV/fm}^3$ and $\epsilon_\mathrm{sym} = 45^{+1.0}_{-3.0}$ ($60^{+8.0}_{-6.0}$) MeV, respectively.
\end{abstract}

\maketitle

\section{\label{sec:intro}Introduction}
At small densities around the nuclear saturation density $n_0$ ($=0.16\ \mathrm{fm}^{-3}$),  the nuclear matter properties are well constrained based on various terrestrial experiments and nuclear theories. The binding energy of nuclear matter at $n_0$ is $B\approx -16$ MeV, while the corresponding incompressibility $K = 240 \pm 20$ MeV~\cite{Shlomo2006_EPJA30-23}, symmetry energy $S = 31.7 \pm 3.2$ MeV and its slope $L = 58.7 \pm 28.1$ MeV~\cite{Li2013_PLB727-276, Oertel2017_RMP89-015007}. In principle, those quantities can be further constrained~\cite{Zhang2020_PRC101-034303, Essick2021_PRL127-192701} if we consider various data from astrophysical observations, heavy-ion collisions, neutron skin thicknesses for $^{208}$Pb in PREX-II~\cite{PREX2021_PRL126-172502} and $^{48}$Ca in CREX~\cite{CREX2022_PRL129-042501}. At larger densities, despite the well established theory of Quantum Chromodynamics (QCD), the properties of strongly interacting matter are nevertheless still veiled in mystery due to its non-perturbative nature, while perturbative calculation is applicable only at densities $n_V\gtrsim 40n_0$~\cite{Fraga2005_PRD71-105014, Fraga2014_ApJ781-L25}. In particular, the uncertainties of nuclear matter properties at densities larger than $2n_0$ are still sizable~\cite{Dutra2012_PRC85_035201, Dutra2014_PRC90-055203, Xia2020_PRD102-023031, LI2020, Hebeler2021_PR890-1}, where the corresponding equation of state (EOS) and matter contents in the core regions of massive compact stars are poorly understood.

Fortunately, astrophysical constraints on the EOSs of dense stellar matter have reached unprecedent accuracy based on various pulsar observations in the multimessenger era. The mass measurements of two-solar-mass pulsars PSR J1614-2230 ($1.928 \pm 0.017\ M_\odot$)~\cite{Demorest2010_Nature467-1081, Fonseca2016_ApJ832-167} and PSR J0348+0432 ($2.01 \pm 0.04\ M_\odot$)~\cite{Antoniadis2013_Science340-1233232} have excluded various soft EOSs. The observations of binary neutron star merger event {GRB} 170817A-{GW}170817-{AT} 2017gfo suggests that the tidal deformability of $1.4 M_{\odot}$ neutron star lies within $70\leq \Lambda_{1.4}\leq 580$ and the corresponding radii $R=11.9\pm1.4$ km~\cite{LVC2018_PRL121-161101}, indicating a soft EOS at small densities. Additionally, the binary compact star merger events GW190814 and GW200210 indicate the existence of massive companion stars with masses $2.6\ M_\odot$ and $2.8\ M_\odot$~\cite{LVC2020_ApJ896-L44, Zhu2022_ApJ928-167}, providing the first observed compact objects in the hypothesized lower mass gap 2.5-5 $M_\odot$~\cite{Yang2020_ApJ901-L34}. If they were neutron stars, then the corresponding EOSs should be much more stiffer than expected. Carrying out pulse-profile modelings using {NICER} and {XMM}-Newton data, the masses and radii of PSR J0030+0451 and PSR J0740+6620 have been measured with similar radii ($\sim$12.4 km) but with large difference in masses ($\sim$1.4 and 2$\ M_\odot$)~\cite{Riley2019_ApJ887-L21, Riley2021_ApJ918-L27, Miller2019_ApJ887-L24, Miller2021_ApJ918-L28}, while the recent measurements of the nearest and brightest millisecond pulsar PSR J0437-4715 ($M=1.418\pm0.037\ M_\odot$ and $R=11.36_{-0.63}^{+0.95}$ km) are more consistent with the constraints derived from gravitational wave measurements of neutron star binary mergers~\cite{Choudhury2024}.

Based on those observational data, we can then constrain the EOSs of neutron star matter at large densities. To fix the properties of dense stellar matter, in this work we adopt the relativistic-mean-field (RMF) models~\cite{Brockmann1977_PLB69-167, Boguta1981_PLB102-93, Mares1989_ZPA333-209, Mares1994_PRC49-2472, Toki1994_PTP92-803, Song2010_IJMPE19-2538, Tanimura2012_PRC85-014306, Wang2013_CTP60-479, Meng2016_RDFNS}, which was shown to give satisfactory description for both finite nuclei~\cite{Reinhard1989_RPP52-439, Ring1996_PPNP37_193-263, Meng2006_PPNP57-470, Paar2007_RPP70-691, Meng2015_JPG42-093101, Chen2021_SCPMA64-282011, Typel1999_NPA656-331, Vretenar1998_PRC57-R1060, Lu2011_PRC84-014328, Wei2020_CPC44-074107, Taninah2020_PLB800-135065} and nuclear matter~\cite{Glendenning2000, Ban2004_PRC69-045805, Weber2007_PPNP59-94, Long2012_PRC85-025806, Sun2012_PRC86-014305, Wang2014_PRC90-055801, Fedoseew2015_PRC91-034307, Gao2017_ApJ849-19}. In particular, by neglecting the derivative terms with respect to the space coordinates, we obtain the properties of uniform nuclear matter using the point-coupling models~\cite{Nikolaus1992_PRC46-1757, Rusnak1997_NPA627-495, Buervenich2002_PRC65-044308, Zhao2010_PRC82-054319}, where nucleons interact with each other only through effective pointlike interactions. In contrast with previous studies, we have assumed density-dependent coupling constants for the effective nucleon-nucleon interactions in the isoscalar-scalar, isoscalar-vector, and isovector-vector channels, where the corresponding values above nuclear saturation densities are constrained according to various pulsar observations.

The purpose of our current study is thus twofold, i.e., constrain the EOSs of neutron star matter and fix the density-dependent coupling constants for RMF models. We then perform a detailed statistical analysis of the density-dependent coupling constants for point-coupling RMF models within a Bayesian approach, where various constraints from astrophysical observations have been considered. At subsaturation densities, we adopt the coupling constants derived from the covariant density functional DD-ME2~\cite{Lalazissis2005_PRC71-024312}, which gives satisfactory description for both nuclear matter and finite nuclei. At densities above the nuclear saturation density, as the constraints from finite nuclei become less significant, we consider six additional free parameters for the coupling constants, which are constrained according neutron star properties. Based on the constrained coupling constants, the nuclear matter properties and the corresponding EOSs are then obtained, where the corresponding energy per nucleon and pressure for symmetric nuclear matter (SNM), pure neutron star matter (PNM), and neutron star matter are presented. The correlations among various nuclear matter properties and coupling constants are investigated as well.

The paper is organized as follows. In Sec.~\ref{sec:the} we present the theoretical framework for the point-coupling RMF model and the Bayesian inference approach. The obtained EOSs and structures of neutron stars are presented in Sec.~\ref{sec:results}, while the constraints and possible correlations among the coupling constants and various nuclear matter properties are investigated. We draw our conclusion in Sec.~\ref{sec:con}

\section{\label{sec:the}Theoretical framework}

\subsection{\label{sec:the_RMF}RMF models}
The Lagrangian density of RMF models for finite nuclei and nuclear matter reads
\begin{eqnarray}
\mathcal{L}
 &=& \bar{\psi}
       \left[  i \gamma^\mu \partial_\mu - \gamma^\mu \left(g_\omega\omega + g_\rho\rho\tau_3\right)- M - g_{\sigma} \sigma \right] \psi
\nonumber \\
 &&\mbox{} + \frac{1}{2}\partial_\mu \sigma \partial^\mu \sigma  - \frac{1}{2}m_\sigma^2 \sigma^2
           - \frac{1}{4} \omega_{\mu\nu}\omega^{\mu\nu} + \frac{1}{2}m_\omega^2 \omega^2
\nonumber \\
 &&\mbox{} - \frac{1}{4} \rho_{\mu\nu}\rho^{\mu\nu} + \frac{1}{2}m_\rho^2 \rho^2 + U(\sigma, \omega),
\label{eq:Lagrange}
\end{eqnarray}
where $\tau_3$ is the 3rd component of isospin and $M$ the nucleon mass. The field tensors $\omega_{\mu\nu}$ and $\rho_{\mu\nu}$ are fixed with
\begin{eqnarray}
\omega_{\mu\nu} &=& \partial_\mu \omega_\nu - \partial_\nu \omega_\mu, \\
\boldsymbol{\rho}_{\mu\nu}
  &=& \partial_\mu \boldsymbol{\rho}_\nu - \partial_\nu \boldsymbol{\rho}_\mu.
\end{eqnarray}
The meson fields $\sigma$, $\omega$, and $\rho$ take mean values and are left with the time components due to time-reversal symmetry, while only the 3rd component of $\rho$ meson remains due to charge conservation.

For uniform nuclear matter, the derivative terms of meson fields vanish, the Lagrangian density in Eq.~(\ref{eq:Lagrange}) can then be reduced to point-coupling versions~\cite{Nikolaus1992_PRC46-1757}, i.e.,
\begin{eqnarray}
 {\cal L} &=& \bar{\psi}\left(i\gamma^\mu\partial_\mu-M\right)\psi
     +\frac{1}{2}\alpha_S\left(\bar\psi\psi\right)^2
     -\frac{1}{2}\alpha_V\left(\bar\psi\gamma^\mu\psi\right)^2 \nonumber\\
     && -\frac{1}{2}\alpha_{TV}\left(\bar\psi\gamma^\mu\tau_3\psi\right)^2.
\label{Eq:LAG}
\end{eqnarray}
The four-fermion coupling constants $\alpha_S$, $\alpha_V$, and $\alpha_{TV}$ are the interaction strengths in the isoscalar-scalar, isoscalar-vector, and isovector-vector channels, which correspond to the exchange of $\sigma$, $\omega$, and $\rho$ mesons in the finite-range RMF models, i.e.,
\begin{equation}
\alpha_S = g_\sigma^2/m_\sigma^2, \ \
\alpha_V = g_\omega^2/m_\omega^2, \ \
\alpha_{TV} = g_\rho^2/m_\rho^2. \label{Eq:mscoup}
\end{equation}

In this work, instead of adopting the nonlinear self couplings $U(\sigma, \omega)$ in Eq.~(\ref{eq:Lagrange}), we consider density-dependent coupling constants to account for in-medium effects. Adopting Euler-Lagrange equation, one obtains the equation of motion for nucleons, i.e.,
\begin{eqnarray}
&&\left[-i\mathbf{\alpha}\cdot \mathbf{\nabla} + \beta\left(M-\alpha_S n_S\right)
      \right] \psi \nonumber \\
 =&&\left(\epsilon - \Sigma^R - \alpha_V n_V - \tau_3 \alpha_{TV} n_{TV} \right) \psi,
\label{Eq:Dirac}
\end{eqnarray}
The rearrangement term $\Sigma^R$ emerges due to the density dependence of coupling constants, which is given by
\begin{equation}
\Sigma^R  = -\frac{1}{2} \frac{\mbox{d} \alpha_S}{\mbox{d} n_V}n_S^2 + \frac{1}{2} \frac{\mbox{d} \alpha_V}{\mbox{d} n_V}n_V^2  + \frac{1}{2} \frac{\mbox{d} \alpha_{TV}}{\mbox{d} n_V}n_{TV}^2.
\end{equation}
The chemical potential of nucleons is then
\begin{equation}
\mu_i =\sqrt{\nu_i^2 + {M^*}^2} + \Sigma^R + \alpha_V n_V + \tau_{3, i} \alpha_{TV} n_{TV}.
\label{Eq:mu_i}
\end{equation}
Here $M^{*}=M-\alpha_S n_S$ is the effective nucleon mass and $\nu_i$ the Fermi momentum, which determines the particle number density $n_i=\nu_i^3/3\pi^2$. The energy density of nuclear matter is given by
\begin{equation}
E_\mathrm{NM} =  E_{\rm{k}} + \frac{1}{2} \alpha_{S}n_{S}^2 +\frac{1}{2} \alpha_{V}n_{V}^2 +\frac{1}{2} \alpha_{TV}n_{TV}^2,
\label{eq:ener_dens}
\end{equation}
where the kinetic energy
\begin{equation}
E_{\rm{k}} =  \frac {{M^*}^4}{8\pi^{2}} \sum_{i=p,n}\left[x_i(2x_i^2+1)\sqrt{x_i^2+1}-\mathrm{arcsh}(x_i) \right]
\end{equation}
with  $x_i\equiv \nu_i/M^*$. The scalar and vector densities are determined by
\begin{eqnarray}
n_{S} &=&  \langle \bar{\psi} \psi \rangle = \frac{{M^*}^3}{2\pi^2} \sum_{i=n,p}\left[x_i \sqrt{x_i^2+1} - \mathrm{arcsh}(x_i)\right],\\
n_V &=& \langle \bar{\psi}\gamma^0 \psi \rangle = n_n + n_p =  \frac{\nu_n^3+\nu_p^3}{3\pi^2}, \\
n_{TV} &=& \langle \bar{\psi}\gamma^0 \tau_3 \psi \rangle = n_n - n_p =\frac{\nu_n^3-\nu_p^3}{3\pi^2}.
\end{eqnarray}
Then pressure is obtained with
\begin{equation}
P_\mathrm{NM} = \sum_{i=p,n}\mu_i n_i - E_\mathrm{NM}.
\end{equation}

To fix the EOSs of uniform neutron star matter, the contribution of electrons and muons needs to be considered. The Lagrangian density for lepton $l$ ($= e,\mu$) is given by
\begin{equation}
  \mathcal{L}_{l} = {\bar{\psi}}_{l}(i\gamma^{\mu}\partial_{\mu} - m_{l})\psi_{l},
\end{equation}
with the lepton masses $m_e=0.511$ MeV and $m_\mu=105.66$ Mev. The lepton densities can be expressed in terms of the corresponding Fermi momenta $n_{l} = \nu_{l}^{3}/3\pi^{2}$, while the chemical potentials are fixed by
\begin{equation}
 \mu_{l} = \sqrt{\nu_{l}^{2} + m_{l}^{2}}.
\end{equation}
To accurately determine the EOSs of uniform neutron star matter, the charge neutrality and $\beta$-equilibrium conditions must be satisfied, i.e.,
\begin{eqnarray}
n_p&=&n_e+n_\mu, \label{Eq:chntral}\\
\mu_{n} &=& \mu_{p} + \mu_{e} = \mu_{p} + \mu_{\mu}.  \label{Eq:beta}
\end{eqnarray}
Note that the contribution of neutrinos are not included since they can leave the system freely. At fixed total baryon number density $n_V$, the particle fractions of protons, neutrons, electrons, and muons are then fixed by fulfilling Eqs.~(\ref{Eq:chntral}) and (\ref{Eq:beta}). The total energy density and pressure for neutron star matter are, respectively, calculated from
\begin{eqnarray}
 E &=& E_\mathrm{NM} +\sum_{l=e,\mu} E_{l}, \label{Eq:E_NS}   \\
 P &=& P_\mathrm{NM} +\sum_{l=e,\mu} P_{l}, \label{Eq:P_NS}
\end{eqnarray}
where the energy density and pressure of leptons are fixed by
\begin{eqnarray}
E_l &=&  \frac{m_{l}^{4}}{8\pi^{2}} \left[x_l(2x_l^2+1)\sqrt{x_l^2+1}-\mathrm{arcsh}(x_l) \right], \\
P_l &=&  \frac{m_{l}^{4}}{24\pi^{2}} \left[x_l(2x_l^2+1)\sqrt{x_l^2+1}+ 3\mathrm{arcsh}(x_l) \right],
\end{eqnarray}
with \(x_{l} = \nu_{l}/m_{l}\) being the dimensionless Fermi momentum.

\subsection{\label{sec:dens}Density-dependent couplings}
To reproduce the correct incompressibility of nuclear matter, the in-medium effects have to be considered. In nonlinear RMF models~\cite{Lalazissis1997_PRC55-540, Long2004_PRC69-034319, Sugahara1994_NPA579-557, Glendenning1991_PRL67-2414, Maruyama2005_PRC72-015802}, the nonlinear self couplings of the mesons are included in the Lagrangian density of Eq.~(\ref{eq:Lagrange}), e.g.,
\begin{equation}
U(\sigma, \omega) = -\frac{1}{3}g_2\sigma^3 - \frac{1}{4}g_3\sigma^4 + \frac{1}{4}c_3\omega^4,  \label{eq:U_NL}
\end{equation}
which effectively introduce density dependent couplings with
\begin{eqnarray}
g_{\sigma}(n_V) &=& g_{\sigma} - \frac{1}{n_S}\frac{\partial U(\sigma, \omega)}{\partial \sigma}, \\
g_{\omega}(n_V) &=& g_{\omega} + \frac{1}{n_V}\frac{\partial U(\sigma, \omega)}{\partial \omega}, \\
g_{\rho}(n_V) &=& g_{\rho}. \label{eq:ddcp_NL}
\end{eqnarray}
Note that the nonlinear point-coupling models give similar effective density dependent couplings~\cite{Nikolaus1992_PRC46-1757, Rusnak1997_NPA627-495, Buervenich2002_PRC65-044308, Zhao2010_PRC82-054319}, which will not be elaborated here.

Alternatively, the in-medium effects can be treated with density dependent coupling constants according to the Typel-Wolter ansatz~\cite{Typel1999_NPA656-331}, such as the covariant density functionals DD-LZ1~\cite{Wei2020_CPC44-074107}, DDME-X~\cite{Taninah2020_PLB800-135065}, PKDD~\cite{Long2004_PRC69-034319}, DD-ME2~\cite{Lalazissis2005_PRC71-024312}, DD2~\cite{Typel2010_PRC81-015803}, and TW99~\cite{Typel1999_NPA656-331}, where
\begin{eqnarray}
g_{\xi}(n_V) &=& g_{\xi} a_{\xi} \frac{1+b_{\xi}(n_V/n_0+d_{\xi})^2}
                          {1+c_{\xi}(n_V/n_0+d_{\xi})^2}, \label{eq:ddcp_TW} \\
g_{\rho}(n_V) &=& g_{\rho} \exp{\left[-a_\rho(n_V/n_0 + b_\rho)\right]}.\label{eq:ddcp_rho}
\end{eqnarray}
Here $\xi=\sigma$, $\omega$ and the baryon number density $n_V = n_p+n_n$.

In this work, instead of adopting those density dependent couplings,  we aim to constrain the couplings at supranuclear densities. In practice, we divide the whole density into three regions, i.e., $n_\mathrm{on}\leq n_V \leq n_0$, $n_0\leq n_V \leq 2 n_0$ and $n_V > 2 n_0$ with $n_\mathrm{on} = 0.1$ fm$^{-3}$. In each density region we adopt the following exponential form for the coupling constants, i.e.,
\begin{eqnarray}
\alpha_S &=&  \frac{\alpha_S'(n_\mathrm{I})^2}{\alpha_S''(n_\mathrm{I})}  \left[\mathrm{e}^{\frac{\alpha_S''(n_\mathrm{I})}{\alpha_S'(n_\mathrm{I})}(n_V-n_\mathrm{I}) }-1 \right]+  \alpha_S(n_\mathrm{I}), \label{eq:non1}\\
\alpha_V &=&  \frac{\alpha_V'(n_\mathrm{I})^2}{\alpha_V''(n_\mathrm{I})}  \left[\mathrm{e}^{ \frac{\alpha_V''(n_\mathrm{I})}{\alpha_V'(n_\mathrm{I})} (n_V-n_\mathrm{I}) }-1 \right]+  \alpha_V(n_\mathrm{I}),\label{eq:non2} \\
\alpha_{TV} &=&  \frac{\alpha_{TV}'(n_\mathrm{I})^2}{\alpha_{TV}''(n_\mathrm{I})}  \left[\mathrm{e}^{ \frac{\alpha_{TV}''(n_\mathrm{I})}{\alpha_{TV}'(n_\mathrm{I})}(n_V-n_\mathrm{I})}-1 \right]+  \alpha_{TV}(n_\mathrm{I}). \label{eq:non3}
\end{eqnarray}
Here $n_\mathrm{I}=n_\mathrm{on}$, $n_0$, and 2$n_0$ correspond to the intersection densities where the coupling constants are matched, i.e., $\alpha_S(n_\mathrm{I}^-)=\alpha_S(n_\mathrm{I}^+)$, $\alpha_V(n_\mathrm{I}^-)=\alpha_V(n_\mathrm{I}^+)$, and $\alpha_{TV}(n_\mathrm{I}^-)=\alpha_{TV}(n_\mathrm{I}^+)$. To avoid discontinuities in the chemical potentials and pressures, we require that the first-order derivatives be continuous, i.e., $\alpha_S'(n_\mathrm{I}^-)=\alpha_S'(n_\mathrm{I}^+)$, $\alpha_V'(n_\mathrm{I}^-)=\alpha_V'(n_\mathrm{I}^+)$, and $\alpha_{TV}'(n_\mathrm{I}^-)=\alpha_{TV}'(n_\mathrm{I}^+)$. The second-order derivatives $\alpha_S''(n_\mathrm{I})$, $\alpha_V''(n_\mathrm{I})$, and $\alpha_{TV}''(n_\mathrm{I})$ are treated as independent parameters to modulate the density-dependence of the coupling constants at densities $n_V \geq n_\mathrm{I}$ until reaching the next matching point. It is important to note that the choice of intersection densities, \( n_\mathrm{I} \), is not unique. Other values could, in principle, be adopted since the precise forms of the density-dependent coupling constants remain uncertain. For instance, shifting \( n_\mathrm{I} = 2n_0 \) to \( 3n_0 \) yields stronger constraints on \( \alpha_S''(n_0) \), \( \alpha_V''(n_0) \), and \( \alpha_{TV}''(n_0) \), which consequently impacts nuclear matter properties. However, we consider \( n_\mathrm{I} = 2n_0 \) to provide more reasonable constraints. Future studies should incorporate a broader range of density segments, with intersection densities selected at random.

\begin{table*}
  \centering
  \caption{\label{table:DDparam} The adopted coefficients for Eqs.~(\ref{eq:non1})-(\ref{eq:non3}) at  subsaturation density $n_\mathrm{on}\leq n_V \leq n_0$, which are fixed based on the covariant density functional DD-ME2~\cite{Lalazissis2005_PRC71-024312}. The corresponding nucleon mass is $M=938.5$ MeV.}
\begin{tabular}{c |c| c}
  \hline
 Zeroth order & First order            & Second order \\
    MeV$^{-2}$ &  MeV$^{-5}$               & MeV$^{-8}$   \\ \hline
   $\alpha_S(n_\mathrm{on})=4.06895\times10^{-4}$  &     $\alpha_S'(n_\mathrm{on})=-1.31629\times10^{-10}$ &    $\alpha_S''(n_\mathrm{on})=1.94990 \times10^{-16}$      \\
   $\alpha_V(n_\mathrm{on})=3.07785\times10^{-4}$  &     $\alpha_V'(n_\mathrm{on})=-1.01883\times10^{-10}$ &    $\alpha_V''(n_\mathrm{on})=1.40885\times10^{-16}$      \\
$\alpha_{TV}(n_\mathrm{on})=3.42884\times10^{-5}$  &  $\alpha_{TV}'(n_\mathrm{on})=-3.31735\times10^{-11}$ & $\alpha_{TV}''(n_\mathrm{on})=3.20949\times10^{-17}$      \\
  \hline
\end{tabular}
\end{table*}

\begin{figure}[ht]
  \centering
  \includegraphics[width=\linewidth]{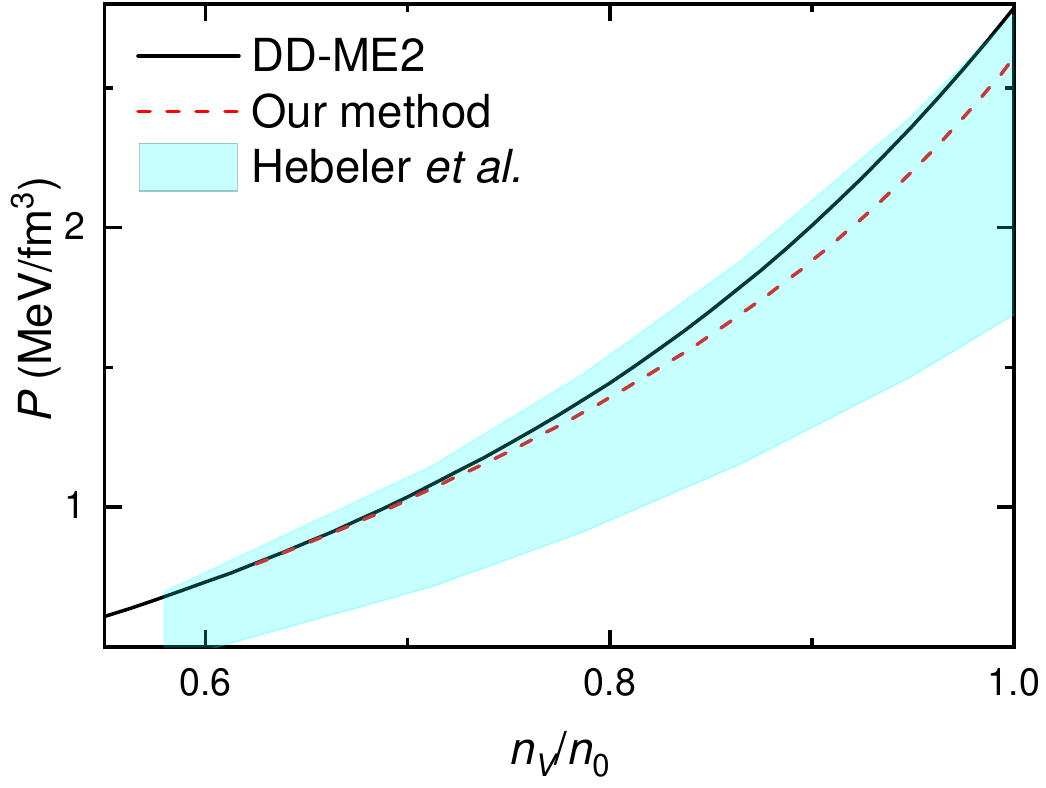}
  \caption{\label{Fig:EOS-DDME2} Pressure of neutron star matter as a function of reduced density, \( n_V/n_0 \), calculated using the covariant density functional DD-ME2~\cite{Lalazissis2005_PRC71-024312} and our density functional with Eqs.~(\ref{eq:non1})-(\ref{eq:non3}) and the parameters in Table~\ref{table:DDparam}. The cyan band represents constraints from chiral effective field theory~\cite{Hebeler2013_ApJ773-11}.}
\end{figure}

Since we are interested in the density dependence of coupling constants and nuclear matter properties at supranuclear densities, at subsaturation densities $n_\mathrm{on}\leq n_V \leq n_0$ with $n_\mathrm{on} = 0.1$ fm$^{-3}$ and $n_0 = 0.16$ fm$^{-3}$ we fix the coefficients based on the covariant density functional DD-ME2~\cite{Lalazissis2005_PRC71-024312}. The corresponding values for the coefficients are listed in Table \ref{table:DDparam}, where the nuclear saturation properties predicted by DD-ME2 are as follows, i.e., the saturation density $n_0=0.152$ fm${}^{-3}$, energy per nucleon $\varepsilon_0=-16.13$ MeV, incompressibility $K=250.8$ MeV, skewness coefficient $J=477$ MeV, symmetry energy $\varepsilon_\mathrm{sym} = 32.3$ MeV, the slope $L=51.2$ MeV and curvature parameter $K_\mathrm{sym} = -87$ MeV of nuclear symmetry energy. Those saturation properties are generally consistent with the state-of-art constraints on nuclear matter, while in principle we could consider $\alpha_{TV}''(n_\mathrm{on})$ as a free parameter since there are still ambiguities in the slope $L(n_\mathrm{on})$ of symmetry energy~\cite{Zhang2020_PRC101-034303, Xie2021_JPG48-025110, PREX2021_PRL126-172502, Essick2021_PRL127-192701, CREX2022_PRL129-042501} even though the symmetry energy at $n_\mathrm{on}$ is well constrained~\cite{Centelles2009_PRL102-122502, Brown2013_PRL111-232502}. Due to the complexity of numerical calculations, we leave this to our future works. In Fig.~\ref{Fig:EOS-DDME2}, we present the pressures of neutron star matter obtained using the covariant density functional DD-ME2~\cite{Lalazissis2005_PRC71-024312} and our density functional, based on Eqs.~(\ref{eq:non1})-(\ref{eq:non3}) and the parameters listed in Table~\ref{table:DDparam}. Our density functional (red dashed curve) closely replicates the equation of state (EOS) predicted by the DD-ME2 functional (black solid curve). Additionally, the constraints from chiral effective field theory, as derived by Hebeler et al.~\cite{Hebeler2013_ApJ773-11}, are shown, indicating that our EOSs are generally in agreement.

\begin{figure}[ht]
  \centering
  \includegraphics[width=\linewidth]{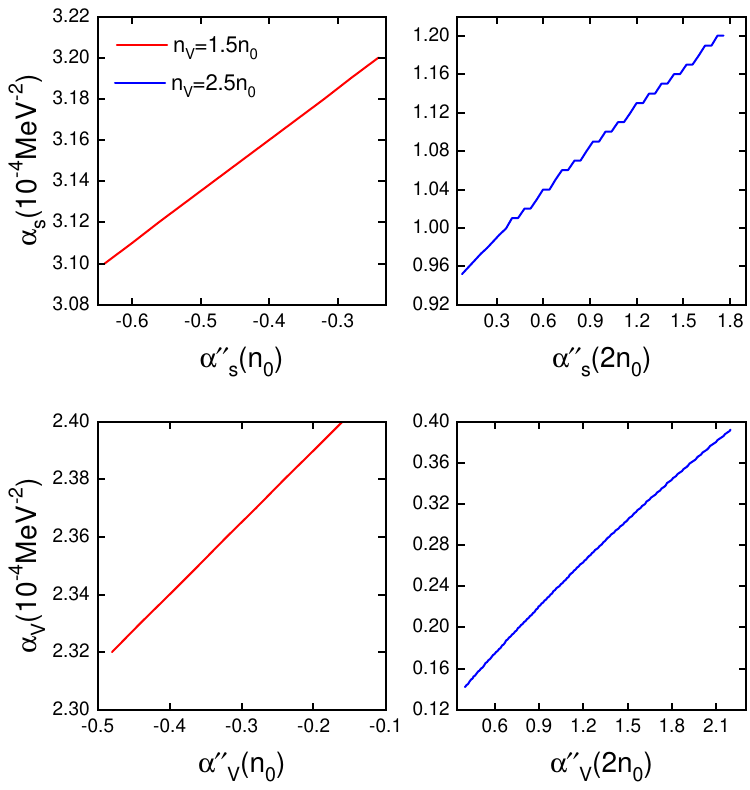}
  \caption{\label{Fig:asddas} Linear correlations between the coupling constants $\alpha_{S,V}(n_V)$ at densities $n_V=$ 1.5$n_0$ and 2.5$n_0$ and the independent parameters $\alpha_{S,V}''(n_0)$ and $\alpha_{S,V}''(2n_0)$.}
\end{figure}

Finally, we are left with six independent parameters that need to be fixed according to astrophysical observations, i.e., $\alpha_S''(n_0)$, $\alpha_V''(n_0)$,  $\alpha_{TV}''(n_0)$, $\alpha_S''(2n_0)$, $\alpha_V''(2n_0)$, and $\alpha_{TV}''(2n_0)$. Note that for neutron star matter at $n_V <n_\mathrm{on}$, we use the unified EOS predicted by DD-ME2 adopting single nucleus approximation~\cite{Xia2022_CTP74-095303}.
As shown in Fig. \ref{Fig:asddas}, a linear correlation is observed between these independent parameters and the coupling constants. As the density increases from 1.5$n_0$ to 2.5$n_0$, a decrease in sensitivity is observed. Specifically, at a density of 1.5$n_0$, the slope of the correlation between $\alpha_S$ and $\alpha_S''(n_0)$ is 0.25, while at 2.5$n_0$, the slope decreases to 0.15.

\subsection{\label{sec:bayes}Bayesian inference approach}
The central concept of the Bayesian inference approach is the Bayes' theorem, which states:
\begin{equation}\label{Bay1}
P({\cal M}|D) = \frac{P(D|{\cal M}) P({\cal M})}{\int P(D|{\cal M}) P({\cal M})d\cal M}.
\end{equation}
In this context, $P({\cal M}|D)$ denotes the posterior probability of the model $\cal M$ given the dataset $D$, which is the central focus of our analysis. The term $P(D|{\cal M})$ refers to the likelihood function, representing the conditional probability that the model $\cal M$ predicts the data $D$ correctly. $P({\cal M})$ is the prior probability of the model $\cal M$ before considering the data. The denominator in Eq. (\ref{Bay1}) serves as a normalization constant, enabling the comparison of multiple models to determine which one provides a more plausible explanation for the experimental data.

The six independent parameters involved in the density-dependent couplings, i.e., $p_{i=1,2,\cdots,6}$ = $\alpha_S''(n_0)$, $\alpha_V''(n_0)$,  $\alpha_{TV}''(n_0)$, $\alpha_S''(2n_0)$, $\alpha_V''(2n_0)$, and $\alpha_{TV}''(2n_0)$, are randomly sampled uniformly within the prior ranges listed in Table \ref{tab-prior}.
For the determination of their minimum and maximum values, we first specify their ranges in the beginning calculations, i.e., the ranges for $\alpha_S''(n_0)$, $\alpha_V''(n_0)$,  $\alpha_{TV}''(n_0)$, $\alpha_S''(2n_0)$ and $\alpha_V''(2n_0)$ are taken as -10 to 10, and $\alpha_{TV}''(2n_0)$ are -10 to 50.
That is, running a wide design range with low statistics, we then adjust the parameter ranges as necessary and re-run with normal precision based on the resulting low-precision posterior distribution, as done in Ref. \cite{Bernhard2018_arxiv}.
Next by substituting them, $p_{i=1,2,\cdots,6}$, into equations (\ref{eq:non1}), (\ref{eq:non2}), and (\ref{eq:non3}), we obtain the density-dependent coupling constants, i.e., $\alpha_S(n_V)$, $\alpha_V(n_V)$, and $\alpha_{TV}(n_V)$. Afterward, these parameters are used as inputs to the RMF model to construct the EOS for neutron stars (NSs) in $\beta$-equilibrium as illustrated in Eqs.~(\ref{Eq:E_NS}) and (\ref{Eq:P_NS}). In the next step, we determine the NS mass-radius sequences by solving the Tolman-Oppenheimer-Volkoff (TOV) equations~\cite{Tolman1934_PNAS20-169, Oppenheimer1939_PR55-374}, i.e.,
\begin{eqnarray}
&&\frac{\mbox{d}P}{\mbox{d}r} = -\frac{G M E}{r^2}   \frac{(1+P/E)(1+4\pi r^3 P/M)} {1-2G M/r},  \label{eq:TOV}\\
&&\frac{\mbox{d}M}{\mbox{d}r} = 4\pi E r^2, \label{eq:m_star}
\end{eqnarray}
where $G=6.707\times 10^{-45}\ \mathrm{MeV}^{-2}$ is the gravity constant.
The resulting theoretical radius $R_{\mathrm{th},j}$ is then used to assess the likelihood of the selected NS EOS reproducing the observed radius $R_{\mathrm{obs},j}$, where $j$ ranges from 1 to 7, as presented in the dataset $D(R_{1,2,\cdots 7})$ in Table \ref{tab-data}. This radius likelihood calculation is given by:
\begin{equation}\label{Likelihood-R}
P_\mathrm{R}[D(R_{1,2,\cdots 7})|{\cal M}(p_{1,2,\cdots 6})]=\prod_{j=1}^{7}\frac{\exp\left[-\frac{(R_{\mathrm{th},j}-R_{\mathrm{obs},j})^{2}}{2\sigma_{\mathrm{obs},j}^{2}}\right]}{\sqrt{2\pi}\sigma_{\mathrm{obs},j}},
\end{equation}
where $\sigma_{\mathrm{obs},j}$ represents the $1\sigma$ error bar associated with observation $j$. For data with different upper and lower 68\% confidence boundaries ($\sigma$), we treat them separately for $R$ values smaller and larger than $R_{\mathrm{obs}}$ to approximate an asymmetric (non-Gaussian) distribution. Additionally, when multiple results from different analyses of the same source, such as GW170817 and PSR J0030+0451, are available, we consider them equally probable. Numerically, we compute their statistical average by randomly selecting one of the results with equal weight to calculate the likelihood function. Note that the recent measurements of PSR J0437-4715 is not included in the study since it generally coincide with the constraints derived from gravitational wave measurements of neutron star binary mergers~\cite{Choudhury2024}.

The total likelihood function can be constructed by multiplying the individual likelihood components. In this work, we employ the following form:
\begin{equation}\label{Likelihood}
  P[D|{\cal M}(p_{1,2,\cdots 6})]=P_{\rm{filter}} \times P_{\rm{mass,max}} \times P_{\rm{R}}.
\end{equation}
Here $P_{\rm{filter}}$ serves as a filter to select EOS parameter sets that meet the criterion where the thermodynamic stability condition (i.e., $dP/dE\geq0$) and the causality condition (i.e., the speed of sound is always less than the speed of light) are satisfied at all densities. The $P_{\rm{mass,max}}$ ensures that each accepted EOS is sufficiently stiff to support the observed maximum mass of neutron stars, $M_{\rm{max}}$. A value of $M_{\rm{max}}=1.97 M_{\odot}$ is adopted in the present work.

A Markov-Chain Monte Carlo (MCMC) method using the Metropolis-Hastings algorithm is employed to simulate the posterior probability density function (PDF) of the model parameters. The PDFs of individual parameters and their pairwise correlations are obtained by integrating over the other parameters. For instance, the PDF for the $i$th parameter $p_i$ is defined as:
\begin{equation}\label{Bay3}
P(p_i|D) = \frac{\int P(D|{\cal M}) dp_1dp_2\cdots dp_{i-1}dp_{i+1}\cdots dp_6}{\int P(D|{\cal M}) P({\cal M})dp_1dp_2\cdots dp_6}.
\end{equation}
Numerically, the initial samples during the burn-in period must be discarded, as the MCMC algorithm does not sample from the equilibrium distribution at the start. Our earlier work suggests that 40,000 burn-in steps are sufficient~\cite{Xie2019_ApJ883-174, Xie2020_ApJ899-4, Xie2023_NST34-91}. Therefore, we discard the first 40,000 steps and utilize the remaining one million steps to compute the posterior PDFs of the six parameters in this analysis.

\begin{table}[htbp]
\centering
\caption{Prior ranges of the parameters used in units of MeV$^{-8}$(values to be multiplied by $10^{-16}$).}\label{tab-prior}
 \begin{tabular}{lccccccc}
  \hline\hline
   Parameters&Lower limit  &Upper limit \\
    \hline 
$\alpha_S''(n_0)$ & $-1.0$ & 0.5 \\
$\alpha_V''(n_0)$ & $-0.8$ & 0.4 \\
$\alpha_{TV}''(n_0)$  & $-0.4$ & 2.0 \\
$\alpha_S''(2n_0)$ & $-0.1$ & 5.5 \\
$\alpha_V''(2n_0)$ & $-0.2$ & 5.0 \\
$\alpha_{TV}''(2n_0)$ & $-0.1$ & 10.0 \\
 \hline
 \end{tabular}
\end{table}

\begin{table}[htbp]
\centering
\caption{Data for the NS radius and mass used in the present work.}\label{tab-data}
 \begin{tabular}{lccccccc}
  \hline\hline
   Mass($\mathrm{M}_{\odot}$)&Radius $R$ (km)  & Source and Reference \\
    \hline\hline 
 1.4 & 11.9$^{+1.4}_{-1.4}$(90\% CFL)&GW170817~\cite{LVC2018_PRL121-161101} \\
1.4 &10.8$^{+2.1}_{-1.6}$ (90\% CFL)&GW170817 \cite{De2018_PRL121-091102} \\
1.4  & 11.7$^{+1.1}_{-1.1}$ (90\% CFL)&QLMXBs \cite{Lattimer2014_EPJA50-40} \\
$1.34_{-0.16}^{+0.15}$ &$12.71_{-1.19}^{+1.14}$ (68\% CFL)&PSR J0030+0451 \cite{Riley2019_ApJ887-L21} \\
$1.44_{-0.14}^{+0.15}$ &$13.0_{-1.0}^{+1.2}$ (68\% CFL)&PSR J0030+0451 \cite{Fonseca2021_ApJ915-L12} \\
$2.08_{-0.07}^{+0.07}$ &$13.7_{-1.5}^{+2.6}$ (68\% CFL)&PSR J0740+6620 \cite{Miller2021_ApJ918-L28} \\
$0.77_{-0.17}^{+0.20}$ &$10.4_{-0.78}^{+0.86}$ (68\% CFL)&HESS J1731-347 \cite{Doroshenko2022_NA6-1444}\\
 \hline
 \end{tabular}
\end{table}

\section{\label{sec:results}Results and discussion}

\begin{figure}[ht]
  \centering
 \includegraphics[width=\linewidth]{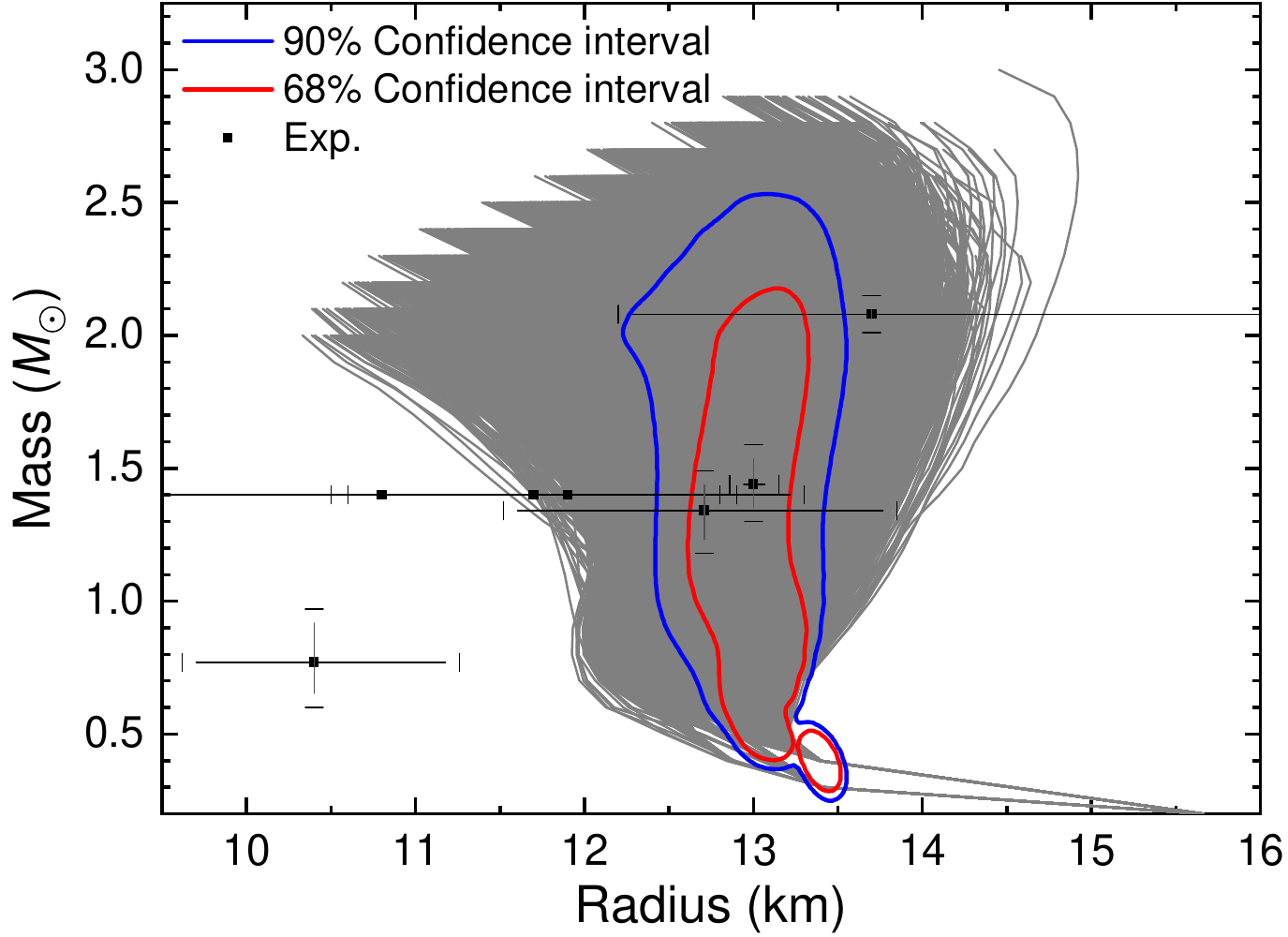}
  \caption{\label{Fig:mr}Mass-radius relations of neutron stars predicted by the RMF models adopting the prior ranges of the parameters listed in Table~\ref{tab-prior}. The posterior 68\%, 90\% credible regions of neutron star masses and radii in the Bayesian analysis are presented as well using the constraints illustrated in Table~\ref{tab-data}.}
\end{figure}

\begin{table}[htbp]
\centering
\caption{Most probable values and their corresponding credible intervals (68\% and 90\%) for the six parameters (values to be multiplied by $10^{-16}$). Their prior ranges are indicated in Table~\ref{tab-prior}.}\label{tab-6parampv}
 \begin{tabular}{lccccccc}
  \hline\hline
   Parameters&68\%  &90\% \\
    \hline 
$\alpha_S''(n_0)$ & $-0.48^{+0.24}_{-0.16}$ & $-0.48^{+0.48}_{-0.2}$ \\
$\alpha_V''(n_0)$ & $-0.36^{+0.20}_{-0.12}$ & $-0.36^{+0.36}_{-0.16}$  \\
$\alpha_{TV}''(n_0)$  & 0.54$^{+0.72}_{-0.24}$ & 0.54$^{+1.2}_{-0.36}$  \\
$\alpha_S''(2n_0)$ & 0.16$^{+1.6}_{-0.08}$ & 0.16$^{+2.64}_{-0.08}$ \\
$\alpha_V''(2n_0)$ & 0.9$^{+1.3}_{-0.5}$ & 0.9$^{+2.1}_{-0.6}$ \\
$\alpha_{TV}''(2n_0)$ & 0.2$^{+6.2}_{-0.0}$ & 0.2$^{+8.8}_{-0.0}$ \\
 \hline
 \end{tabular}
\end{table}

With the EOSs of neutron star matter fixed by Eqs.~(\ref{Eq:E_NS}) and (\ref{Eq:P_NS}) in the framework of RMF models, in Fig.~\ref{Fig:mr} we present the prior mass-radius relations of neutron stars obtained by solving the TOV equations in Eqs.~(\ref{eq:TOV}) and (\ref{eq:m_star}), where the six independent parameters responsible for the density dependent coupling constants are randomly sampled uniformly within the prior ranges specified by Table~\ref{tab-prior}. Those parameters are then constrained according to the astrophysical observations illustrated in Table~\ref{tab-data}, where the most probable values and their corresponding credible intervals are listed in Table~\ref{tab-6parampv}.

The constrained masses and radii of neutron stars in the 68\% and 90\% credible regions are then presented in Fig.~\ref{Fig:mr}. Evidently, the radii of neutron stars predicted by RMF models lie at the upper end of the radius constraints derived from the binary neutron star merger event GW170817~\cite{LVC2018_PRL121-161101, De2018_PRL121-091102} and pulse profile modeling of PSR J0437-4715~\cite{Choudhury2024}. Meanwhile, there is basically no overlap between the probability distributions for the masses and radii of neutron stars predicted by current RMF models and the central compact object (CCO) within the supernova remnant HESS J1731-347~\cite{Doroshenko2022_NA6-1444}. In such cases, the CCO at HESS J1731-347 might be strange stars or other exotic objects~\cite{Sagun2023_ApJ958-49}. Note that beside HESS J1731-347, there might be other compact objects with unusually small masses and radii, e.g., 4U 1746-37~\cite{Li2015_ApJ798-56}. Finally, the 90\% credible interval for the maximum mass of neutron stars in this work is 2.0$^{+0.65}_{-0.0} \mathrm{M}_\odot$,
in contrast with the previous investigations based on RMF models~\cite{Malik2022_ApJ930-17}, we find that it is possible for neutron stars to reach 2.5 $M_\odot$ while meeting all the current observations based on our current theoretical framework, which could be important for us to understand the nature of the secondary objects in the  binary compact star merger events GW190814 and GW200210~\cite{LVC2020_ApJ896-L44, Zhu2022_ApJ928-167} and filling the hypothesized lower mass gap 2.5-5 $M_\odot$~\cite{Yang2020_ApJ901-L34}.

In Table \ref{tab-r14cs}, we present the 90\% confidence intervals for several observational quantities related to neutron stars inferred in this work, including the radii, tidal deformability, central energy density, and central speed of sound of neutron stars with masses 1.4 and 2.0 times the solar mass, respectively. The method used in this study for calculating neutron-star tidal deformability is consistent with that in Ref. \cite{Fattoyev2013_PRC87-015806,xie2023cpc}. These results indicate that the nuclear EOSs obtained in this study are relatively stiff, consistent with previous findings derived using the relativistic mean-field theory. This stiffness results in larger neutron star radii, tidal deformability, and central speed of sound, while yielding a lower central energy density.
Our calculated radii for neutron stars are similar to those reported by the NICER collaboration but are slightly larger than the values published by the LIGO/Virgo collaboration as listed in Table \ref{tab-data}. Regarding tidal deformability, our results are higher than those in Refs. \cite{Fujimoto2020prd} ($320\pm120$) and \cite{Kumar2019prd} (196$^{+92}_{-63}$) but remain within the upper bound ($\Lambda_{1.4}\leq 800$) deduced from observations of the GW170817 event\cite{LVC2017_PRL119-161101}. Using the universal relations among the maximum mass and radius of neutron stars and their central density and pressure, a central energy density of 901$^{+214}_{-287}$ MeV/fm$^{3}$ are derived\cite{Cai2023apj}. Another range of 904$^{+329}_{-327}$ MeV/fm$^{3}$ are obtained by the Bayesian analysis based on neutron star mass and radius data from GW170817\cite{Brandes2023prd}. Although our result is relatively lower, the range of 320 to 1180 MeV/fm$^{3}$ is broadly consistent with these findings.
\begin{table}[htbp]
\centering
\caption{Most probable values and the corresponding 90\% credible intervals for the radius $R$, and tidal deformability $\Lambda$, the energy density $\varepsilon_c$ and the speed of sound $v_{s}^{c}$ at the center of neutron stars for neutron stars with masses of 1.4 and 2.0 solar masses.}\label{tab-r14cs}
 \begin{tabular}{lccccccc}
  \hline\hline
   Mass($\mathrm{M}_{\odot}$)&$R$ (km)  &$\Lambda$ &$\varepsilon_c$(MeV/fm$^{3}$)& $v_{s}^{c}$\\
    \hline\hline 
 1.4 & 13$^{+0.3}_{-0.5}$&605$^{+150}_{-165}$ &340$^{+100}_{-30}$ & 0.7$^{+0.2}_{-0.1}$ \\
2.0 &13.3$^{+0.3}_{-1.3}$&75$^{+15}_{-50}$ &420$^{+440}_{-50}$ & 0.8$^{+0.05}_{-0.4}$\\
 \hline
 \end{tabular}
\end{table}
\begin{figure*}[ht]
  \centering
  \includegraphics[width=0.8\linewidth]{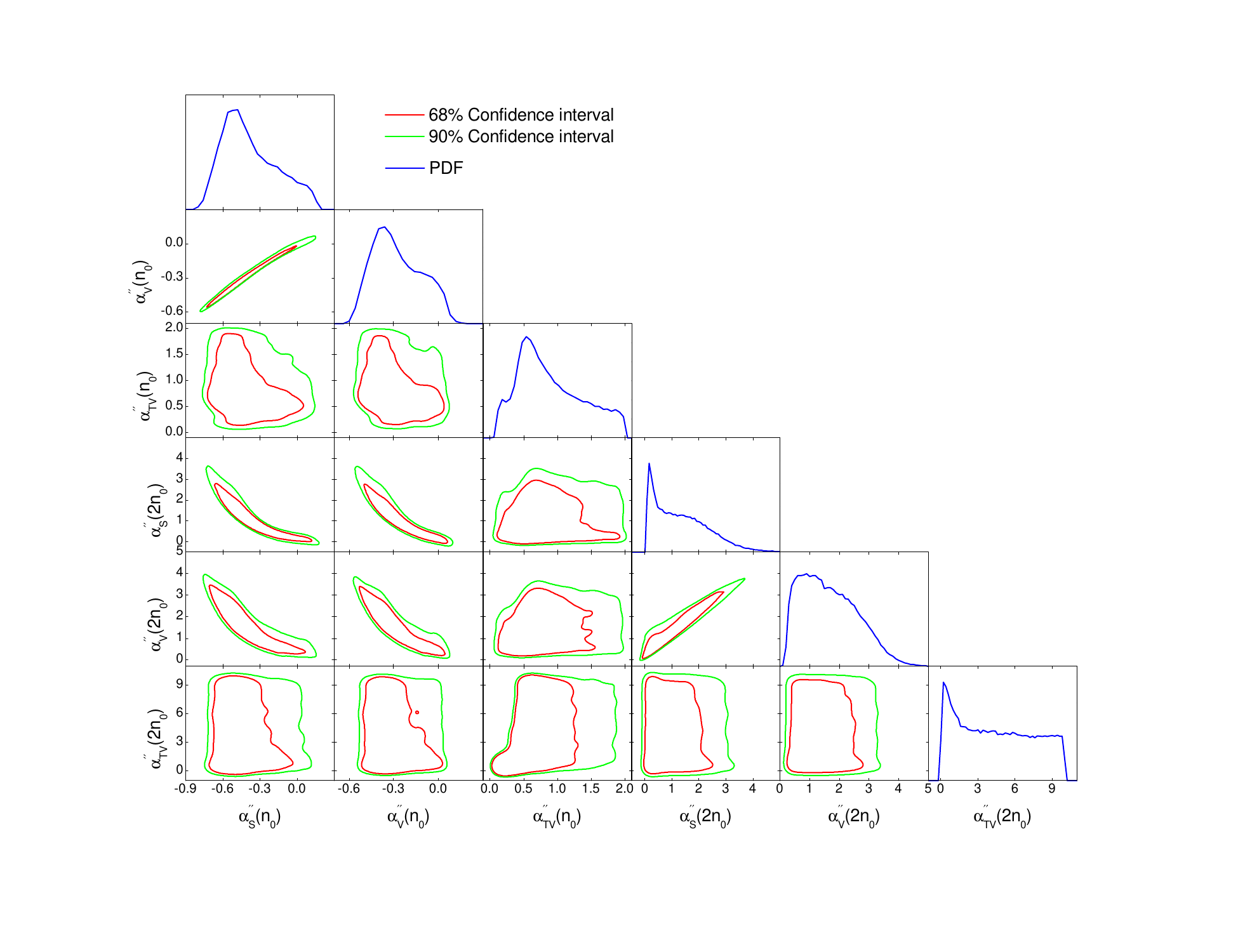}
  \caption{Posterior probability distribution functions of the six parameters (shown by blue curves) and their correlations (shown by red (green) curves for the 68\% (90\%) credible regions) inferred from the Bayesian analysis of the NS radius dataset listed in Table \ref{tab-data}. }\label{Fig:6para}
\end{figure*}

In Fig.~\ref{Fig:6para} the posterior PDFs of the six parameters and their correlations at 68\% and 90\% credible levels inferred from the Bayesian analysis are presented. The obtained probability distributions generally have a single peak structure, suggesting that the parameters are well constrained according to the astrophysical observations used in the present work. The corresponding peak values and the 68\% and 90\% credible intervals are indicated in Table~\ref{tab-6parampv}.
It is found that the peak values fixed at the saturation density, $\alpha_S''(n_0)$ and $\alpha_V''(n_0)$, are negative, while those fixed at twice the saturation density, $\alpha_S''(2n_0)$ and $\alpha_V''(2n_0)$, become positive, with $\alpha_V''(2n_0)$ taking a much larger value. The credible intervals widen at $2n_0$, indicating that the uncertainties becomes sizable at higher densities. Note that a linear correlation between those parameters and the coupling constants were observed in Fig.~\ref{Fig:asddas}, where the correlation are weakened as density increases. For the isovector channel, both $\alpha_{TV}''(n_0)$ and $\alpha_{TV}''(2n_0)$ are positive, with $\alpha_{TV}''(2n_0)$ being smaller. This suggests that the variations of coupling constant for $\alpha_{TV}$ is minimal at higher densities. Generally speaking, the upper limits of $\alpha_V''(2n_0)$ and $\alpha_{TV}''(2n_0)$ are substantial, allowing for stiffer EOSs that enable neutron stars to reach masses exceeding $2.5M_\odot$, thereby approaching the lower mass gap of 2.5-5 $M_\odot$~\cite{Yang2020_ApJ901-L34}. This is mainly attributed to the more independent parameters adopted for density dependent coupling constant, in contrast to previous investigations~\cite{Malik2022_ApJ930-17}. It is noted that the parameter $\alpha_{TV}''(2n_0)$ is not well constrained based on the currently used neutron star observational data. This is primarily because, as density increases, the variation in the nuclear symmetry energy is relatively small, while the EOSs for symmetric nuclear matter changes more significantly. It is therefore less well constrained since $\alpha_{TV}''(2n_0)$ is directly related to the symmetry energy as expressed in Eq. (\ref{Eq:sym}).

In the marginalized posterior distributions shown in Fig.~\ref{Fig:6para}, an elliptical shape indicates correlations between each pair of parameters, whereas a circular or rectangular shape suggests the absence of such correlations. At the same density, such as at $n_0$ and 2$n_0$, the parameters $\alpha_S''$ and $\alpha_V''$ exhibit positive linear correlations. This is because, to ensure the properties at saturation density (using the density functional DD-ME2 in this study, which provides well-determined properties at saturation density), the coupling constants $\alpha_S$ and $\alpha_V$ should be positively linearly correlated. According to Fig. \ref{Fig:asddas}, the parameters $\alpha_S''$ and $\alpha_V''$ are proportional to these coupling constants, leading them to also exhibit a similar positive linear correlation as that of $\alpha_S$ and $\alpha_V$. At higher densities (such as 2$n_0$), this linear correlation weakens. This weakening is attributed to the increased uncertainty in the coupling constants $\alpha_S$ and $\alpha_V$ at 2$n_0$, which in turn reduces the correlation between the associated parameters $\alpha_S''(2n_0)$ and $\alpha_V''(2n_0)$. Interestingly, the correlations between parameters at different densities $\alpha_S''(n_0)/\alpha_V''(n_0)$-$\alpha_S''(2n_0)/\alpha_V''(2n_0)$ display a negative relationship, where the velocity of sound as a function density are altered. In the isovector channel, $\alpha_{TV}''$ is almost uncorrelated with the other parameters.

\begin{figure}[ht]
  \centering
   \includegraphics[width=0.8\linewidth]{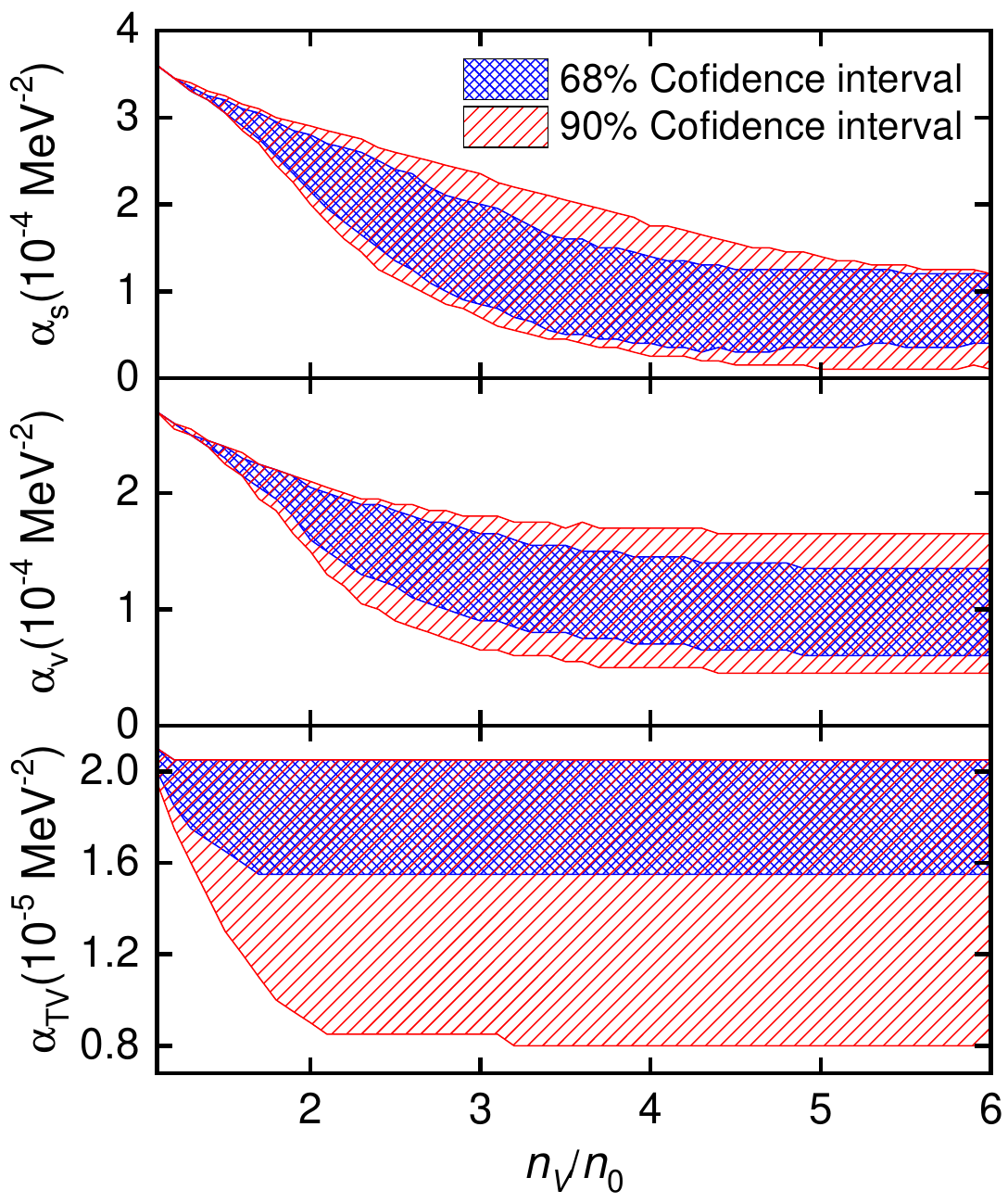}
  \caption{The 68\% and 90\% credible intervals for the three coupling constants as functions of the reduced density.}\label{Fig:asavatvrho}
\end{figure}

Similarly, we can also obtain the credible intervals of the three coupling constants corresponding to the six independent parameters as a function of density, which are presented in Fig. \ref{Fig:asavatvrho}.
Note that at densities below the nuclear saturation density $n_0$, as illustrated in Table~\ref{table:DDparam}, the coupling constants are fixed based on the covariant density functional DD-ME2~\cite{Lalazissis2005_PRC71-024312}. The coupling constants thus start with same values at $n_V=n_0$, where the uncertainties increase with density and are constrained according to the pulsar observations.
It is found that the coupling constants generally decrease and approach to small positive values at sufficiently high densities. In these cases, the uncertainties do not significantly increase with density.
 This however may change if we introduce more free parameters by considering additional intersection densities $n_\mathrm{I}$ in Eqs.~(\ref{eq:non1}-\ref{eq:non3}). Notably, the coupling constant for the isovector-vector channel $\alpha_{TV}$ initially decreases with density and then stabilizes at higher densities. This behavior indicates that the coupling constant $\alpha_{TV}$ is not sensitive to the current neutron star observational data at high densities. In conjunction with Eq. (\ref{Eq:sym}), this finding suggests that the symmetry energy of nuclear matter exhibits minimal variation at high densities. Consequently, it can be inferred that in the high-density region, the contribution of the symmetric nuclear matter equation of state to the internal pressure of neutron stars is more significant than that of the symmetry energy component.
\begin{table*}[htbp]
\centering
\caption{Most probable values and their 68\% and 90\% credible intervals for the coupling constants $\alpha_S(n_V)$, $\alpha_V(n_V)$, and $\alpha_{TV}(n_V)$; the energy per nucleon for symmetric nuclear matter $\varepsilon_0(n_V)$ and pure neutron matter $\varepsilon_N(n_V)$; the symmetry energy $\varepsilon_{\rm{sym}}(n_V)$; the pressure for neutron star matter $P(n_V)$ and pure neutron matter $P_N(n_V)$; and the slope $L(n_V)$, and curvature $K_{\rm{sym}}(n_V)$ of the symmetry energy. Each column corresponds to the values at 1.5$n_0$, 2.5$n_0$, and 5$n_0$, respectively, with 68\% and 90\% credible intervals provided for each parameter.}\label{tab-MP1}
 \begin{tabular}{ll|ccccccc}
  \hline\hline
  \multicolumn{2}{c|}{Quantities} &1.5$n_0$ &2.5$n_0$ &5$n_0$ \\
  & &68\%, 90\% &68\%, 90\% &68\%, 90\% \\
  \hline
 $\alpha_S(n_V)$ & $[10^{-4}\ \mathrm{MeV}^{-2}]$ &3.08$^{+0.11}_{-0.03}, 3.08^{+0.16}_{-0.05}$ &$1.63^{+0.79}_{-0.29}, 1.63^{+1.0}_{-0.5}$&$1.21^{+0.05}_{-0.87}, 1.21^{+0.2}_{-0.16}$\\
 $\alpha_V(n_V)$ & $[10^{-4}\ \mathrm{MeV}^{-2}]$ &2.29$^{+0.1}_{-0.0}, 2.29^{+0.12}_{-0.03}$ &$1.29^{+0.57}_{-0.13}, 1.29^{+0.64}_{-0.38}$&$1.13^{+0.25}_{-0.52}, 1.13^{+0.54}_{-0.69}$   \\
 $\alpha_{TV}(n_V)$ & $[10^{-5}\ \mathrm{MeV}^{-2}]$ &2.05$^{+0.0}_{-0.4}, 2.05^{+0.0}_{-0.75}$ &$2.05^{+0.0}_{-0.5}, 2.05^{+0.0}_{-1.2}$&$2.05^{+0.0}_{-0.5}, 2.05^{+0.0}_{-1.25}$ \\ \hline
 $\varepsilon_0(n_V)$ & $[\mathrm{MeV}]$  &$-10.4^{+0.6}_{-0.4}, -10.4^{+1.2}_{-1.2}$ &$23.0^{+5.6}_{-12.6}, 23.0^{+11.4}_{-18.8}$&$142.2^{+77.2}_{-22.6}, 142.2^{+153.8}_{-28.8}$\\
 $P(n_V)$ & $[\mathrm{MeV/fm}^3]$  &5.4$^{+1.0}_{-0.4}, 5.4^{+2.0}_{-1.4}$ &$62.0^{+13.4}_{-23.0}, 62.0^{+20.0}_{-36.8}$&$164.6^{+143.8}_{-51.6}, 164.6^{+297.4}_{-60.0}$\\
 $\varepsilon_\mathrm{sym}(n_V)$ & $[\mathrm{MeV}]$  &46.0$^{+0.5}_{-3.5}, 46.0^{+1.0}_{-6.0}$ &$58.0^{+10.0}_{-4.0}, 58.0^{+15.0}_{-10.5}$&$95.0^{+16.5}_{-10.5}, 95.0^{+25.0}_{-25.0}$\\
 $\varepsilon_N(n_V)$ & [MeV] & 37.0$^{+0.5}_{-3.0}$, 37.0$^{+1.0}_{-5.5}$ &$83.0^{+6.0}_{-9.5}, 83.0^{+12.5}_{-14.5}$&$241.5^{+73.5}_{-27.0}, 241.5^{+149.5}_{-36.0}$\\
 $P_N(n_V)$ &$[\mathrm{MeV/fm}^3]$ &$14.0^{+1.5}_{-1.5}$, $14.0^{+2.0}_{-3.5}$,&$75.0^{+9.0}_{-23.5}, 75.0^{+19.5}_{-32.0}$&$251.5^{+110.0}_{-82.5}, 251.5^{+257.5}_{-95.0}$\\
 $L(n_V)$ & $[\mathrm{MeV}]$  &111$^{+9.0}_{-27}, 111^{+18}_{-48}$ &$95^{+35}_{-15}, 95^{+60}_{-30}$&$250^{+30}_{-80}, 250^{+50}_{-130}$\\
 $K_\mathrm{sym}(n_V)$ & $[\mathrm{MeV}]$  &$-200^{+120}_{-120}, -200^{+210}_{-180}$ &$-315^{+525}_{-45}, -315^{+750}_{-45}$&$160^{+10}_{-800}, 160^{+10}_{-1070}$\\
  \hline
 \end{tabular}
\end{table*}

Based on the constrained coupling constants, we can then estimate the properties of nuclear matter and neutron star matter in the framework of RMF models introduced here. The energy per nucleon for SNM can be expressed in terms of the energy density of nuclear matter $E_\mathrm{NM}$ in Eq.~(\ref{eq:ener_dens}) and the mass of nucleons $M$ as
\begin{equation}
\varepsilon_0(n_V) =\left.E_\mathrm{NM}\right|_{n_p=n_n}/n_V-M. \label{Eq:E0}
\end{equation}
For pure neutron matter, the density of protons vanishes with $n_p=0$, then $n_{S}= \frac{{M^*}^3}{2\pi^2} \left[x_n \sqrt{x_n^2+1} - \mathrm{arcsh}(x_n)\right]$ and $n_V = n_{TV}  = n_n =\nu_n^3/3\pi^2$. The corresponding energy density and pressure can then be fixed by
\begin{eqnarray}
E_{N} &=& \frac {{M^*}^4}{8\pi^{2}}\left[x_n(2x_n^2+1)\sqrt{x_n^2+1}-\mathrm{arcsh}(x_n) \right] \nonumber \\
 && + \frac{1}{2} \alpha_{S}n_{S}^2 +\frac{1}{2} (\alpha_{V}+\alpha_{TV})n_{n}^2,\\
P_N &=& \frac{{M^*}^{4}}{24\pi^{2}} \left[x_n(2x_n^2+1)\sqrt{x_n^2+1}+ 3\mathrm{arcsh}(x_n) \right]\nonumber \\
 && \Sigma^R n_n - \frac{1}{2} \alpha_{S}n_{S}^2 +\frac{1}{2} (\alpha_{V}+\alpha_{TV})n_{n}^2,
\end{eqnarray}
where $x_n = \nu_n/{M^*}$ is the dimensionless Fermi momentum for neutrons. The energy per nucleon for pure neutron matter is thus calculated by $\varepsilon_N=E_{N}/n_n$.

The symmetry energy is formulated as
\begin{equation}
 \varepsilon_{\rm{sym}}(n_V) = \frac{\nu^2}{6E_F^*}+\frac{1}{2}\alpha_{TV} n_V, \label{Eq:sym}
\end{equation}
where $n_p=n_n=\nu^3/3\pi^2$ and $E_F^* = \sqrt{\nu^2 + {M^*}^2}$.

\begin{figure}[ht]
  \centering
   \includegraphics[width=0.8\linewidth]{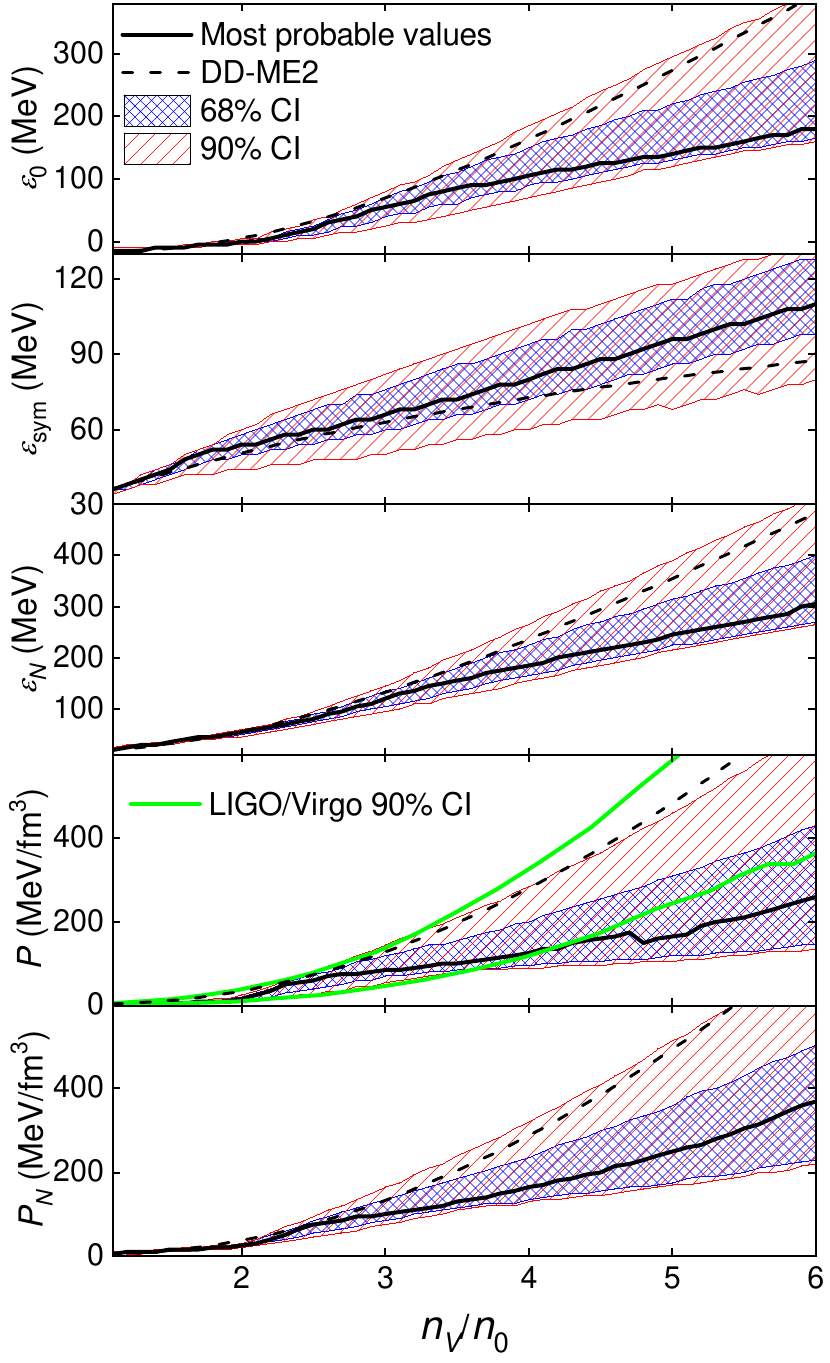}
  \caption{Most probable values and their 68\% and 90\% credible intervals (CIs) for the energy per nucleon of symmetric nuclear matter, $\varepsilon_0(n_V)$, pure neutron matter, $\varepsilon_N(n_V)$, the symmetry energy, $\varepsilon_{\rm{sym}}(n_V)$, and the pressure for neutron star matter, $P(n_V)$, and pure neutron matter, $P_N(n_V)$, as functions of the reduced density. The corresponding results predicted by the covariant density functional DD-ME2~\cite{Lalazissis2005_PRC71-024312} are indicated by the dashed curves as well. For comparison, the pressure as a function of density in Ref. \cite{Abbott2018prl} is included.
}\label{Fig:e0enpBands}
\end{figure}

Shown in Fig. \ref{Fig:e0enpBands} are the 68\% and 90\% credible bands of the energy per nucleon of symmetric nuclear matter, $\varepsilon_0(n_V)$, pure neutron matter, $\varepsilon_N(n_V)$, the symmetry energy, $\varepsilon_{\rm{sym}}(n_V)$, and the pressure for neutron star matter, $P(n_V)$, and pure neutron matter, $P_N(n_V)$, as functions of the reduced density $n_V/n_0$. We also include the results for $P(n_V)$ reported by the LIGO/Virgo Collaboration in the figure~\cite{Abbott2018prl}. It is found that all quantities increase monotonically with density, and their associated uncertainties also grow. The pressure calculated in this work is slightly lower than the values reported by the LIGO/Virgo Collaboration in the high-density region. Once again, we observe that as density increases, the growth of the symmetry energy is much smaller compared to the increase in the symmetric nuclear matter component.

\begin{figure}[ht]
  \centering
   \includegraphics[width=0.8\linewidth]{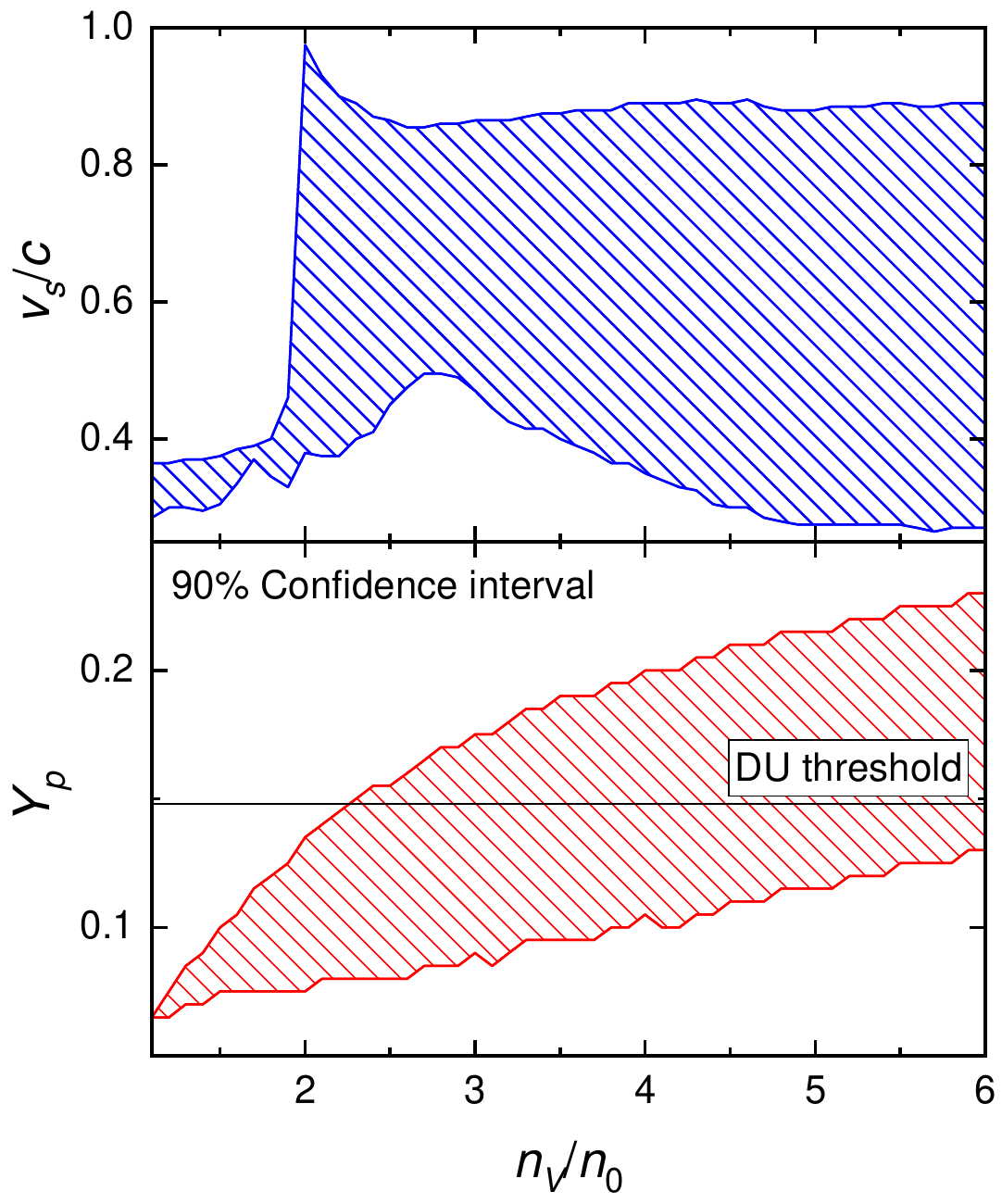}
  \caption{The speed of sound (\(v_s\)) and proton fraction (\(Y_p\)) of neutron star matter are presented within their 90\% credible intervals. The horizontal line in the lower panel indicates the threshold proton fraction of 14.8\%, above which the direct Urca process occurs.
}
  \label{Fig:vsyp}
\end{figure}

The constrained speed of sound and proton fraction in neutron star matter are shown in Fig.~\ref{Fig:vsyp}, where the speed of sound $v_s$ is fixed by
\begin{equation}
  v_s =c \sqrt{\frac{\mbox{d}P}{\mbox{d}E}}.
\end{equation}
To meet the causality condition, we have included a filter $P_\mathrm{filter}$ to select EOS parameters in the likelihood function (\ref{Likelihood}) so that $v_s$ never exceeds $c$.
Interestingly, a peak in \( v_s \) appears as a function of density, despite the absence of any new degrees of freedom. This observation aligns with previous studies~\cite{Jin2022_PLB829-137121} and may suggest a crossover from hadronic to quark matter~\cite{Annala2023_NC14-8451}, similar to how the hadron resonance gas model describes the transition from hadronic matter to quark-gluon plasma at high temperatures~\cite{Karsch2003_PLB571-67, Andronic2018_Nature561-321}. The proton fraction in neutron star matter generally increases with density, influencing neutron star cooling mechanisms. Once the proton fraction surpasses the critical threshold of approximately 14.8\%~\cite{Lattimer1991_PRL66-2701}, the Direct Urca (DU) processes, \( n \rightarrow p + e^- + \bar{\nu}_e \) and \( p + e^- \rightarrow n + \nu_e \), activate, rapidly cooling neutron stars by emitting thermal neutrinos. Notably, the proton fraction may exceed 14.8\% at \( n_V \gtrsim 2n_0 \), indicating an early onset of DU processes in lower-mass neutron stars (\( \lesssim 1 M_\odot \)). Conversely, there are cases where the DU processes do not occur within the density range. Observations of neutron star cooling could therefore constrain the EOS, particularly $\alpha_{TV}(n_V)$, if the star's mass is known.

The slope $L$ and curvature $K_\mathrm{sym}$ of the symmetry energy are obtained with
\begin{eqnarray}
L &=& 3n_V\left(\frac{\partial \varepsilon_{\rm{sym}}}{\partial n_V}\right) \nonumber \\
  &=& \frac{\nu^2}{3E_F^* }
  -\frac{\nu^4}{6{E_F^*}^3}\left(1+\frac{2M^*\nu}{\pi^2}\frac{\partial M^*}{\partial n_V}\right) \nonumber \\
 && \mathrm{} + \frac{3}{2}  n_V \left(\alpha_{TV} +\alpha_{TV}' n_V \right),
\end{eqnarray}
and
\begin{equation}
K_{\rm sym} = 9n_V^2\left(\frac{\partial^2\varepsilon_{\rm{sym}}}{\partial n_V^2}\right) = 9n_{V}^{2}\left( \xi_{1} + \xi_{2} + \xi_{3} + \xi_{4} \right)
\end{equation}
with
\begin{eqnarray}
   \xi_{1} &=& - \frac{\pi^{2}}{12{E_{F}^{*}}^{3} \nu}\left( \frac{\pi^{2}}{\nu} + 2M^{*}\frac{\partial M^{*}}{\partial n_{V}} \right) - \frac{\pi^{4}}{12E_{F}^{*}\nu^{4}}\nonumber, \\
   \xi_{2} &=& - \left\lbrack \frac{\pi^{4}}{24{E_{F}^{*}}^{3}\nu^{2}} - \frac{\pi^{2}\nu}{8{E_{F}^{*}}^{5}}\left( \frac{\pi^{2}}{\nu} + 2M^{*}\frac{\partial M^{*}}{\partial n_{V}} \right) \right\rbrack \nonumber \\
   &&{} \times   \left( 1 + \frac{2M^{*}\nu}{\pi^{2}}\frac{\partial M^{*}}{\partial n_{V}} \right)\nonumber , \\
   \xi_{3} &=& - \frac{\nu \pi^{2}}{12{E_{F}^{*}}^{3}}\left\lbrack \frac{M^{*}}{\nu^{2}}\frac{\partial M^{*}}{\partial n_{V}} + \frac{2\nu M^{*}}{\pi^{2}}\frac{\partial^{2}M^{*}}{\partial n_{V}^{2}}\right. \nonumber  \\
   &&{} \left. + \frac{2\nu}{\pi^{2}}\left( \frac{\partial M^{*}}{\partial n_{V}} \right)^{2} \right\rbrack\nonumber, \\
   \xi_{4} &=& {\alpha}'_{TV} + \frac{1}{2}n_{V}\alpha''_{TV}\nonumber.
\end{eqnarray}
Their posterior PDFs and correlations at 1.5$n_0$ and 2.5$n_0$ are indicated in Fig. \ref{Fig:LKsymCor}, and the most probable values as well as the 68\% and 90\% credible intervals are listed in Table \ref{tab-MP1}.
It is worth noting that Ref. \cite{Li2021universe} summarizes the current constraints on the parameters $L$ and $K_{\text{sym}}$ at $n_0$ as well as the symmetry energy of nuclear matter at $2n_0$. Although our results cannot be directly compared with these constraints, we find that our symmetry energy at $2.5n_0$ appears to be consistent with the result from Ref. \cite{Li2021universe}, where the symmetry energy at $2n_0$ is constrained to $51 \pm 13$ MeV at 68\% confidence level.
At 1.5$n_0$, there is little correlation between $L$ and $K_\mathrm{sym}$. However, as the density increases to 2.5$n_0$, they exhibit an inverse correlation. During the transition from $n_0$ to 2.5$n_0$, $L$ initially increases and then decreases, while $K_\mathrm{sym}$ decreases monotonically (note that at $n_0$ $L$ is 51.2 MeV and $K_\mathrm{sym}$ is $-87$ MeV in the DD-ME2 density functional used in this analysis).

\begin{figure}[ht]
  \centering
  \includegraphics[width=\linewidth]{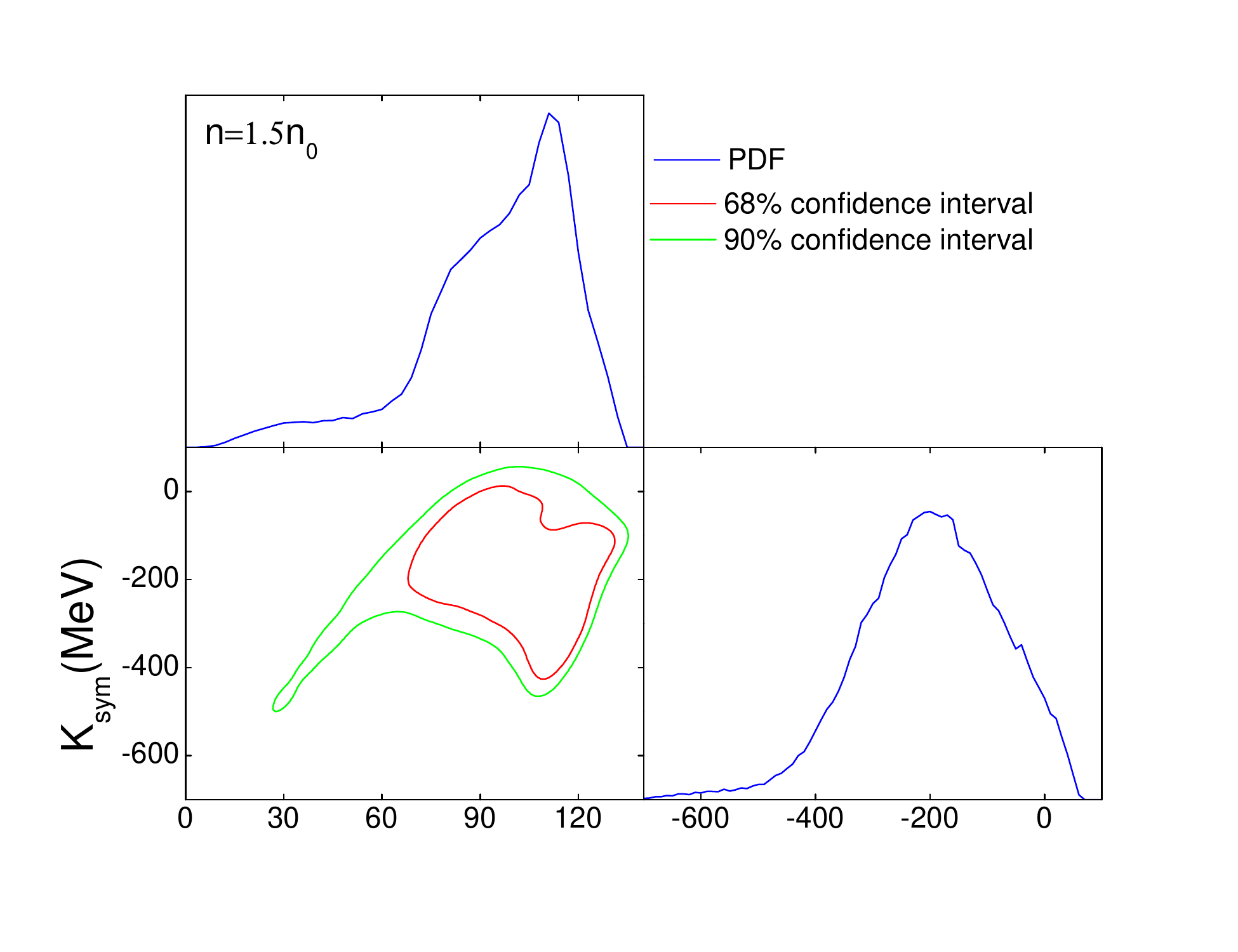}
  \includegraphics[width=\linewidth]{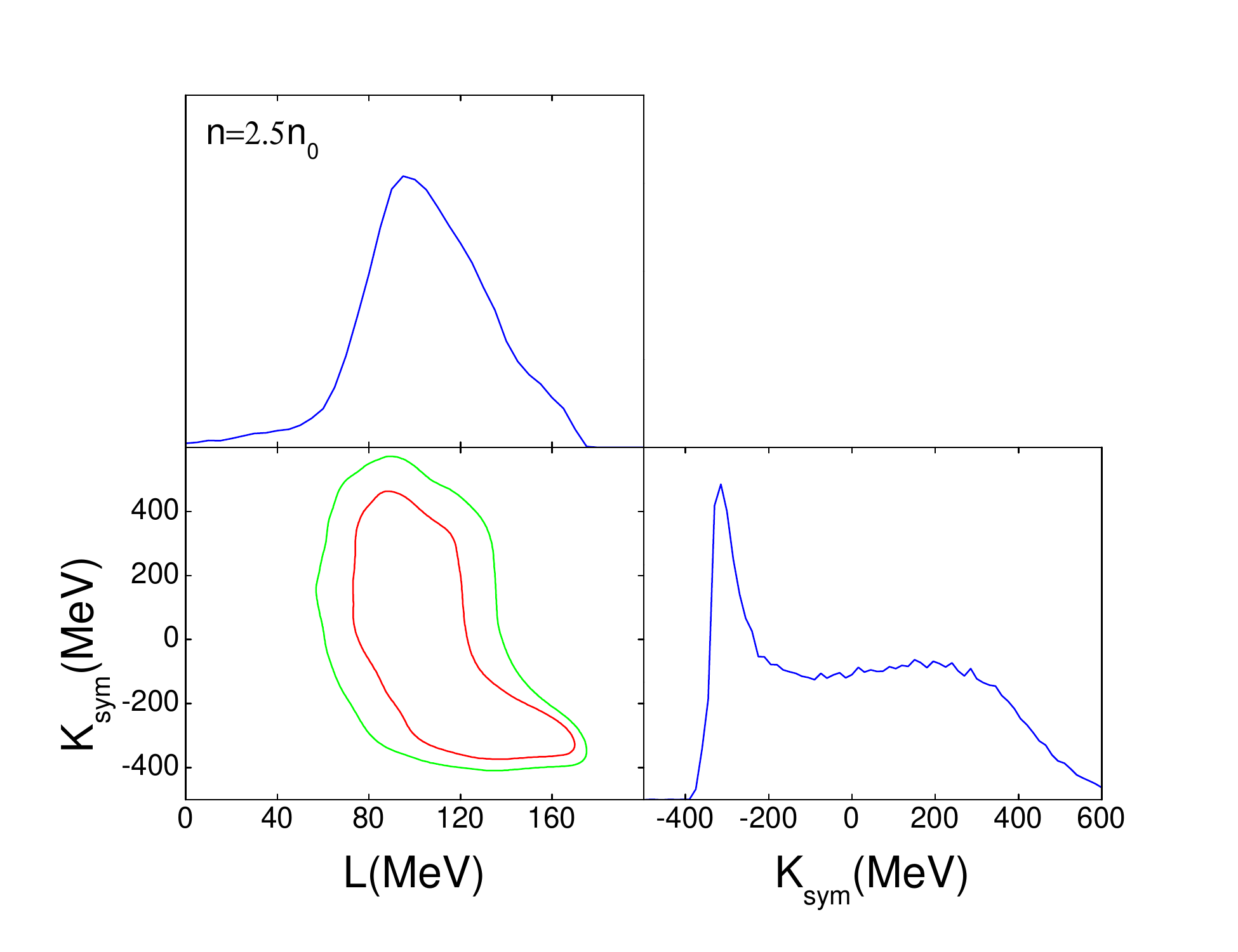}
  \caption{\label{Fig:LKsymCor} Posterior probability distribution functions of the slope and curvature of the symmetry energy at  1.5$n_0$, 2.5$n_0$, where their correlations inferred from the Bayesian analysis of the NS radius dataset listed in Table \ref{tab-data} are indicated as well.}\label{Fig:LKsymCor}
\end{figure}

\begin{figure*}[ht]
  \centering
  \includegraphics[width=\linewidth]{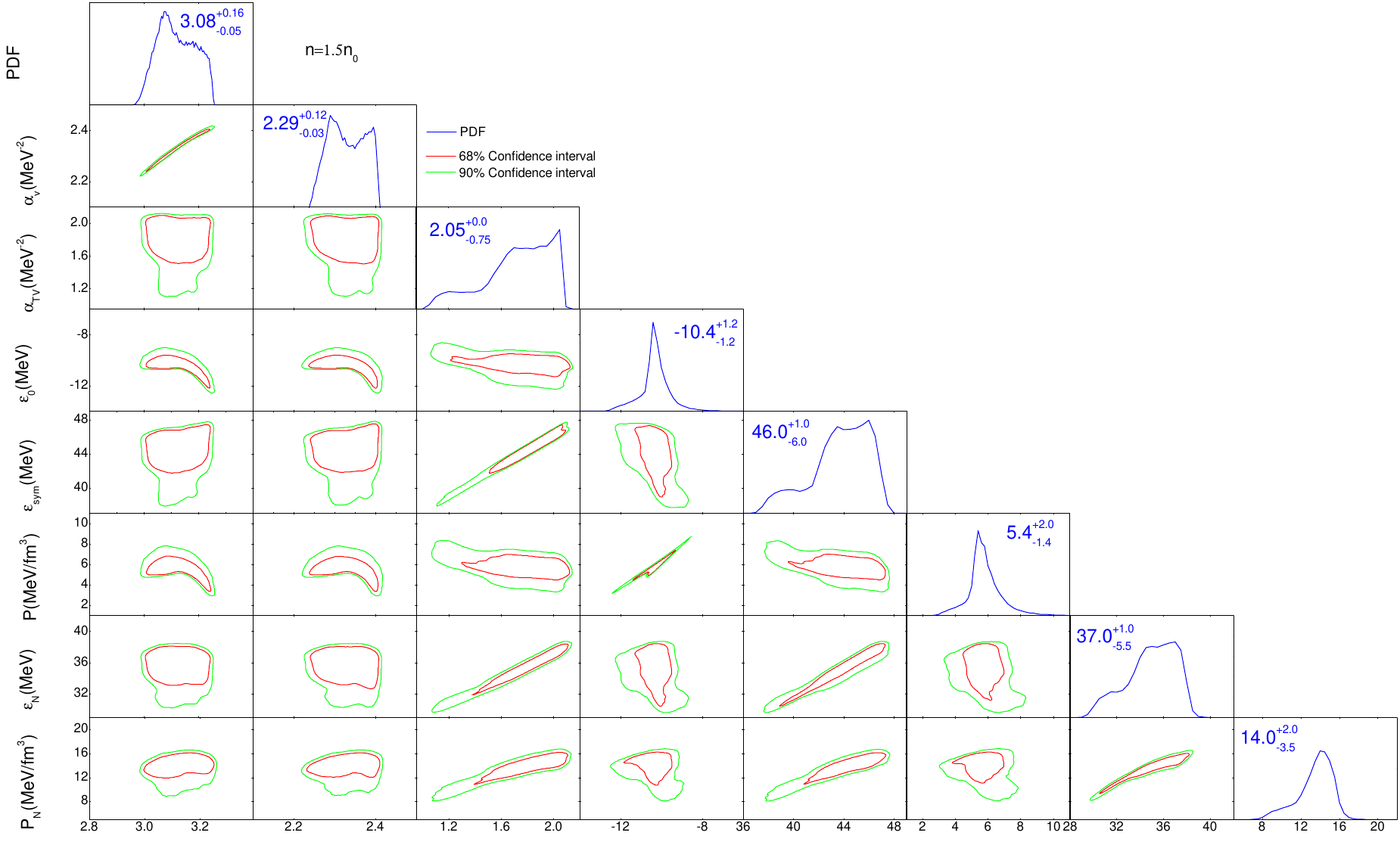}
  \includegraphics[width=\linewidth]{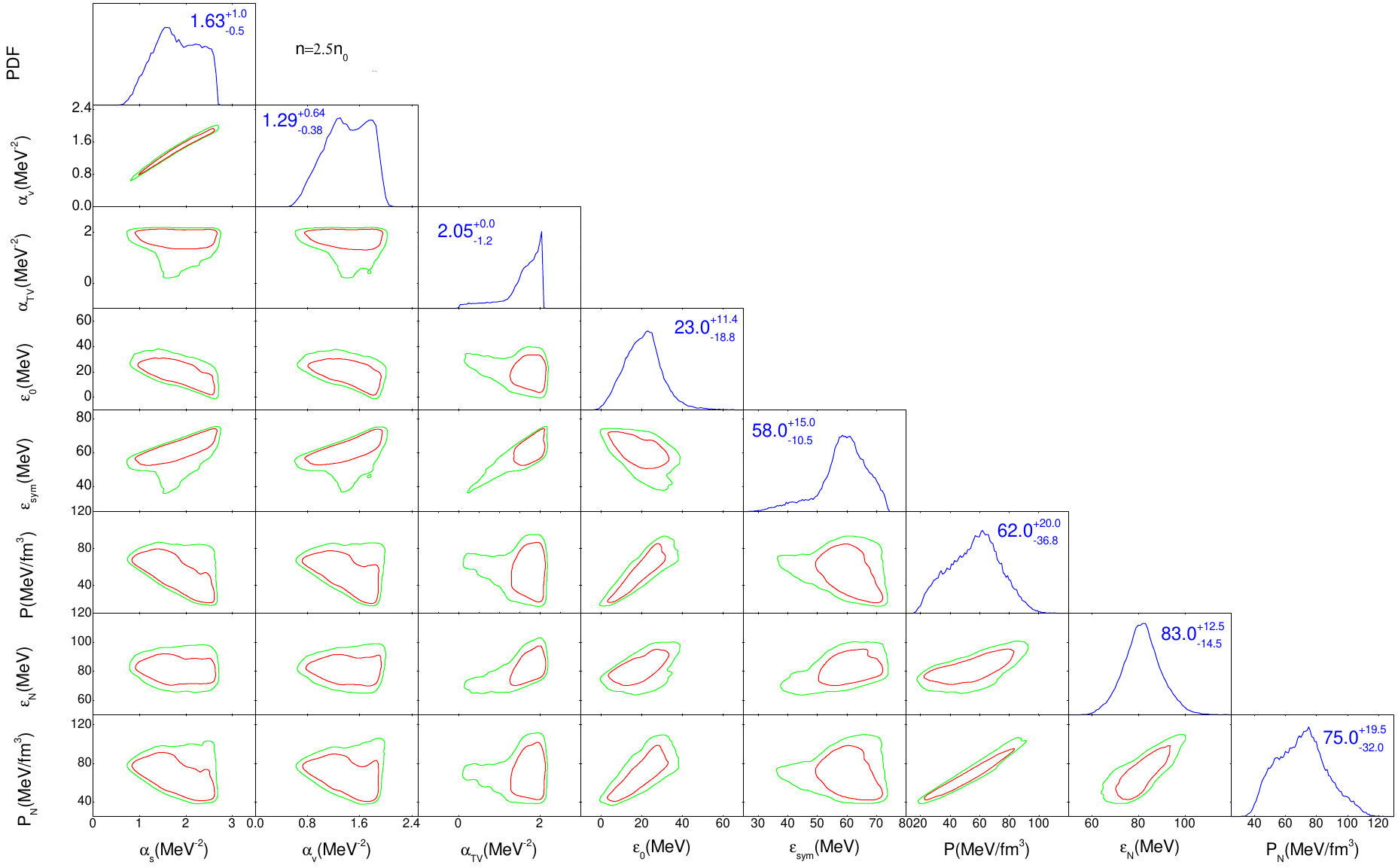}
  \caption{\label{Fig:8paraCor} Posterior probability distribution functions of the coupling constants $\alpha_S$, $\alpha_V$, and $\alpha_{TV}$, the energy per nucleon for symmetric nuclear matter $\varepsilon_0$, the symmetry energy $\varepsilon_{\rm{sym}}$, the pressure $P$, the energy per nucleon and pressure for pure neutron matter $\varepsilon_N$, $P_N$ (shown by blue curves) and their correlations (shown by red (green) curves for the 68\% (90\%) credible regions) at  1.5$n_0$ and 2.5$n_0$ inferred from the Bayesian analysis of the NS radius dataset listed in Table \ref{tab-data}. The scale values of $\alpha_S$ and $\alpha_V$ ($\alpha_{TV}$) have already been  divided by $10^{-4}$ ($10^{-5}$). The most probable values and their 90\% credible intervals are indicated on the diagonal panels.}
\end{figure*}

\begin{figure*}[ht]
  \centering
  \includegraphics[width=\linewidth]{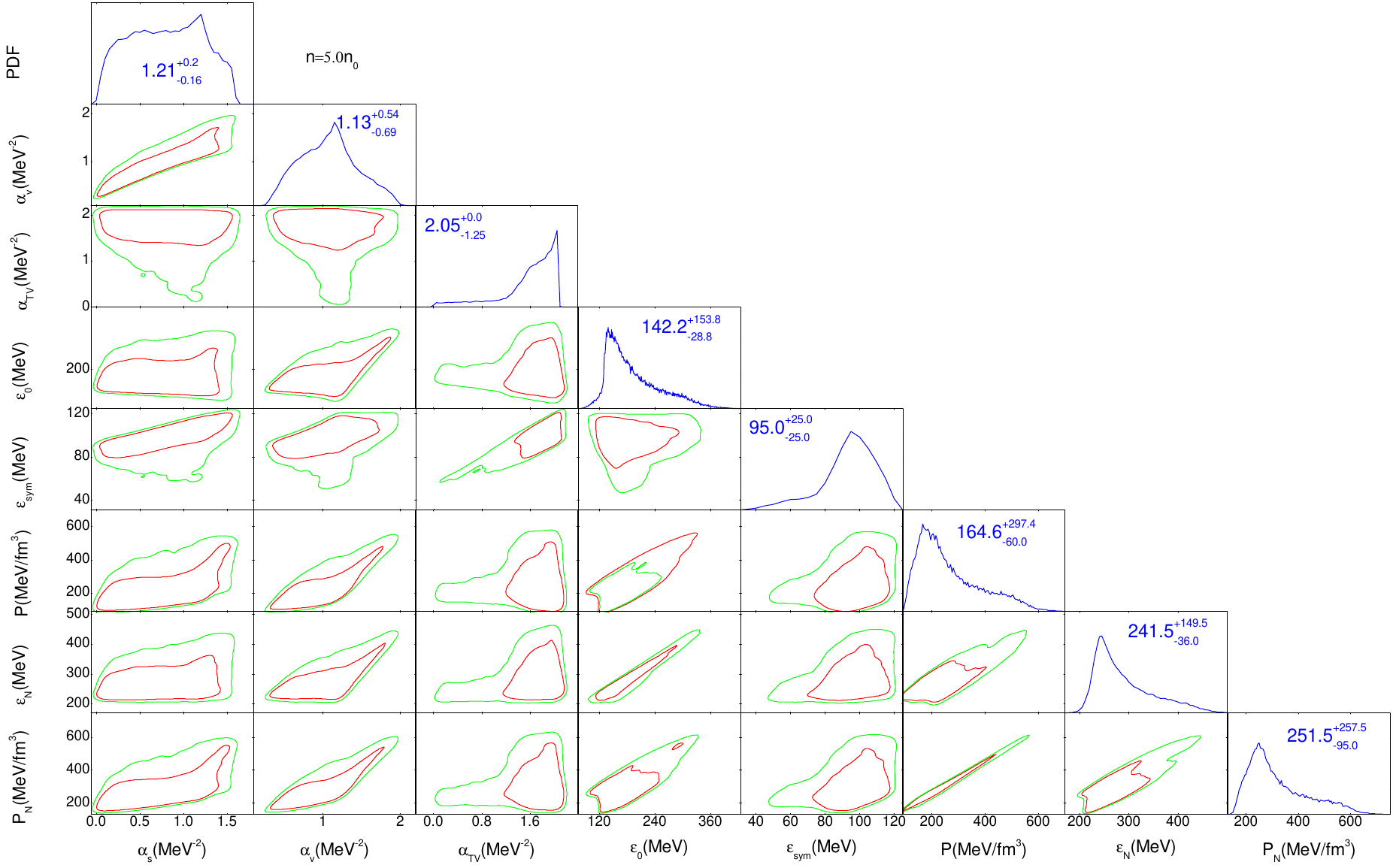}
  \caption{\label{Fig:8paraCor-rho50} Same as Fig.~\ref{Fig:8paraCor} but at density 5$n_0$.}
\end{figure*}

In Figs. \ref{Fig:8paraCor} and \ref{Fig:8paraCor-rho50}, we present the posterior PDFs (shown by blue curves) of the coupling constants $\alpha_S$, $\alpha_V$, and $\alpha_{TV}$, the energy per nucleon for symmetric nuclear matter $\varepsilon_0$, the symmetry energy $\varepsilon_{\rm{sym}}$, the pressure $P$, the energy per nucleon and pressure for pure neutron matter $\varepsilon_N$, $P_N$ and their correlations (shown by red/green curves for the 68\%/90\% credible regions) at 1.5$n_0$ and 2.5$n_0$ inferred from the Bayesian analysis of the NS radius dataset listed in Table \ref{tab-data}. The situation at higher densities, such as 5$n_0$, is shown in Fig. \ref{Fig:8paraCor-rho50}. The obtained most probable values and the 68\%, 90\% credible intervals (CIs) are summarized in Table \ref{tab-MP1}. We can derive the following interesting findings:
\begin{itemize}
\item
Compared to the prior distributions of these quantities, the observational data of neutron star radius used in this analysis provide effective constraints on these parameters, especially at 1.5 $n_0$, as shown in Fig. \ref{Fig:e0enpBands}.
\item
As discussed previously, the coupling constants $\alpha_S$ and $\alpha_V$ exhibit a strong positive linear correlation, which diminishes as the density increases. In the low-density region, a linear relationship between $\alpha_S$ and $\alpha_V$ is established to achieve specific properties at saturation density. However, as the density increases, the uncertainties in $\alpha_S$ and $\alpha_V$ grow, leading to a weakening of this linear correlation.
\item
The coupling constants $\alpha_S$ and $\alpha_V$ are negatively correlated with $\varepsilon_0$ and $P$, and this correlation also weakens with increasing density. This behavior is closely related to the internal mechanisms of the RMF model. In the RMF model,  $\alpha_S$ and $\alpha_V$ correspond to the strength of the attractive and repulsive forces in nucleon interactions, respectively. Typically, when $\alpha_S$ is held constant, the $\alpha_V$ is inversely proportional to $\varepsilon_0$, and vise versa if we take $\alpha_V$ as a constant. As the density increases, the growing uncertainties in $\alpha_S$, $\alpha_V$, $\varepsilon_0$ and $P$ result in a weakening of this correlation.
\item
The parameter $\alpha_{TV}$ shows little to no correlation with $\alpha_S$ and $\alpha_V$. It is positively correlated with the symmetry energy, and this correlation weakens as the density increases. In the low-density region (e.g., at 1.5$n_0$), $\alpha_{TV}$ is positively correlated with both $\varepsilon_N$ and $P_N$, indicating that in the low-density regime, $\varepsilon_{\rm{sym}}$ plays a more crucial role in the equation of state of pure neutron matter than $\varepsilon_0$. However, this relationship disappears in the high-density region due to the weakening constraints on $\alpha_{TV}$ at high densities. The relationship between $\alpha_{TV}$ and $\varepsilon_{\rm{sym}}$, as described by Eq. (\ref{Eq:sym}), helps to explain the observed correlations between $\varepsilon_{\rm{sym}}$ and other quantities.
\item
 $\varepsilon_0$ is positively correlated with $P$, and this correlation diminishes with increasing density. This implies that, in the low-density region, the internal pressure of neutron stars is mainly contributed by the equation of state of symmetric nuclear matter. As the density increases, contributions from other components (such as the symmetry energy and leptons) become more significant. In the low-density region, $\varepsilon_0$ shows little to no correlation with $\varepsilon_N$ and $P_N$, but as the density increases, $\varepsilon_0$ becomes positively correlated with them. This suggests that, in the low-density region, the primary contributor to the energy per nucleon or internal pressure of pure neutron matter is not $\varepsilon_0$, but its contribution increases as the density rises.
 \item
 In the low-density region, there is little correlation between $P$ and $P_N$, but this correlation strengthens as the density increases.
\end{itemize}

\section{\label{sec:con}Conclusion}
In this work we investigate the density dependence of coupling constants in RMF models and the corresponding nuclear matter properties at supranuclear densities, where a Bayesian approach is adopted using various astrophysical constraints of neutron stars. The effective nucleon interactions in the isoscalar-scalar, isoscalar-vector, and isovector-vector channels are considered, where six independent parameters are adopted to modulate the density dependent coupling constants at densities exceeding the nuclear saturation density $n_0$. At subsaturation densities, instead, we match the coupling constants with that of DD-ME2, where the corresponding unified EOS is adopted for neutron star crusts.

The constrained coupling constants are found decreasing with density and approaching to small positive numbers at large enough densities. Such a behavior qualitatively agrees with previous density dependent behaviors in RMF models. At $1\sigma$ level, the constrained coupling constants at densities $1.5n_0$, $2.5n_0$, and $5n_0$ are respectively given by
\begin{eqnarray}
   10^4\alpha_S &=& 3.1^{+0.1}_{-0.05}, 1.55^{+0.85}_{-0.2}, 1.2^{+0.05}_{-0.85} \ \mathrm{MeV}^{-2}, \nonumber \\
   10^4\alpha_V &=&  2.3^{+0.1}_{-0.0}, 1.3^{+0.55}_{-0.1}, 1.15^{+0.2}_{-0.55} \ \mathrm{MeV}^{-2}, \nonumber \\
   10^5\alpha_{TV}  &=&  2.05^{+0}_{-0.4}, 2.05^{+0}_{-0.5}, 2.05^{+0}_{-0.5} \ \mathrm{MeV}^{-2}. \nonumber
\end{eqnarray}
The nuclear matter properties are constrained as well, where the pressure of neutron star matter and symmetry energy at densities $1.5n_0$, $2.5n_0$, and $5n_0$ at $1\sigma$ level are $P=10^{+0.0}_{-0.0}$, $60^{+15}_{-20}$,  $165^{+145}_{-45}$ $\mathrm{MeV/fm}^3$ and $\varepsilon_\mathrm{sym} = 45^{+1.0}_{-3.0}, 60^{+8.0}_{-6.0}, 95^{+17}_{-9}$ MeV, respectively.

The correlations among the coupling constants and various nuclear matter properties are investigated. The strong positive linear correlation between coupling constants $\alpha_S$ and $\alpha_V$ diminishes with increasing density, becoming weaker due to growing uncertainties at higher densities. The negative correlation between $\alpha_S$ and $\alpha_V$ with $\varepsilon_0$ and $P$ also weakens as density increases, due to growing uncertainties in these parameters within the RMF model. In contrast, the parameter $\alpha_{TV}$ shows minimal correlation with $\alpha_{S}$ and $\alpha_{V}$ but is positively correlated with symmetry energy at low densities, although this correlation weakens and eventually disappears at higher densities due to decreasing constraints on $\alpha_{TV}$. Additionally, $\varepsilon_0$ is positively correlated with $P$ in low densities, but this correlation weakens as density increases, indicating that while $\varepsilon_0$ contributes less to neutron star pressure at low densities, its contribution grows with increasing density. Finally, while there is little correlation between $P$ and $P_N$ in the low-density region, this correlation strengthens as density increases.

\section*{ACKNOWLEDGMENTS}
C.J.X. is supported by the National SKA Program of China (Grant Nos. 2020SKA0120300) and the National Natural Science Foundation of China (Grant No. 12275234). W.J.X. is supported by the Shanxi Provincial Foundation for Returned Overseas Scholars under Grant No. 20220037, the Natural Science Foundation of Shanxi Province under Grant No. 20210302123085, the Open Project of Guangxi Key Laboratory of Nuclear Physics and Nuclear Technology under Grant No. NLK2023-03, and the Central Government Guidance Funds for Local Scientific and Technological Development, China, under Grant No. Guike ZY22096024.

\newpage

%

\end{document}